\documentclass[twocolumn,superscriptaddress,letter]{revtex4}
\usepackage{amssymb}
\usepackage{color}
\usepackage{graphicx}
\usepackage{dcolumn}
\usepackage{subfigure}
\usepackage{bm}
\usepackage[header,title,page,titletoc]{appendix}
\usepackage{amsmath}
\usepackage{bbm}
\usepackage{amsfonts}%
\usepackage[dvipdfm,bookmarks,colorlinks=false]{hyperref}
\newcommand{\beq}{\begin{eqnarray}}
\newcommand{\eeq}{\end{eqnarray}}
\newcommand{\be}{\begin{equation}}
\newcommand{\ee}{\end{equation}}
\newcommand{\bw}{\begin{widetext}}
\newcommand{\ew}{\end{widetext}}
\newcommand{\ba}{\begin{array}}
\newcommand{\ea}{\end{array}}
\newcommand{\bk}{\mathbf{k}}
\newcommand{\br}{\mathbf{r}}

\newcommand{\bphi} {{\mbox{\boldmath$\phi$}}}
\newcommand{\bn}{\mathbf{n}}

\newcommand{\bP}{\mathbf{P}}

\newcommand{\bv}{\mathbf{v}}

\begin{document}
\title{Phase shift formulas for baryon-baryon scattering in elongated boxes}

\author{Ning Li}
\email[Corresponding author. Email: ]{lining@xatu.edu.cn}

\author{Ya-Jie Wu}
\affiliation{
School of Science, Xi'an Technological University, Xi'an  710032, P.~R.~China
}

\author{Zhan-Wei Liu}
\affiliation{
School of Physical Science and Technology, Lanzhou University, Lanzhou 730000, China
}

\begin{abstract}
We have established the relations between the baryon-baryon scattering phase shifts and the two-particle energy spectrum in the elongated box. We have studied the cases with both the periodic boundary condition and twisted boundary condition in the center of mass frame. The framework is also extended to the system of nonzero total momentum with periodic boundary condition in the moving frame. This will be helpful to extract the phase shifts in the continuum from lattice QCD data using asymmetric volumes.

%
%

\end{abstract}



\maketitle
\section{INTRODUCTION}
The phase shift for strong elastic scattering of two
hadrons is encoded in the basic knowledge on the strong interaction
which is one of the four elementary interactions in nature.
The phase shift $\delta_l$ is related to the phase difference between
the outgoing and the ingoing $l$ wave outside
the interaction range, and parameterizes the complicated form of this interaction. The scattering length can
be extracted from its effective range expansion.
The knowledge from the phase shift also serves to study the resonance, such as in reference \cite{Wilson:2014cna,Wilson:2015dqa,Dudek:2016cru,Aoki:2007rd,Feng:2010es,Lang:2011mn,Feng:2014gba,Briceno:2016mjc}. Among these references, the authors determined the masses, the widths of
the resonances $K^{\ast}(892)$,
$\rho$, and $a_{0}$
by calculating the scattering phase shifts of
$\pi{K}$, $\eta{K}$ coupled channels, $\pi\pi$, $K\bar{K}$ coupled channels, and $\pi\eta$, $K\bar{K}$ coupled channels, and etc.
A phase shift crossing through $\pi/2$ is
often indicative of a resonance, while the sharpness of the rise allows to
determine the property of resonance according to Breit-Wigner type functional form.

In the aspect of particle and nuclear physics, unraveling the origin of baryon forces based on the
quantum chromodynamics (QCD)  is one of the most challenging issues.
The precise informations on nuclear and hyperon forces serve as the key ingredients to calculate properties of
nuclei, dense matter and the structure of neutron stars \cite{Sun:2014aya,Liu:2015ktc,Liu:2016uzk,Liu:2016wxq,Kiratidis:2016hda}.
According to the baryon-baryon scattering phase shifts that are measured in experiments, people established the relationships on realistic nuclear forces
and the fundamental theories such as QCD.
However, scattering experiments with
hyperons are very difficult because of their short lives.
Then, hyperon forces suffer from large uncertainties. Under these circumstances, it is
most desirable to carry out the first-principles calculations of baryon forces by lattice QCD.
By measuring appropriate correlation functions,
energy eigenvalues of two-particle states in a finite
box can be obtained. L\"uscher found out a relation, now commonly known as
L\"uscher's formula, which relates the energy of two-particle state
in a finite box of size $L$, i.e., $E(L)$, to the elastic scattering phaseshift
$\delta(E(L))$ of the two particles in the
continuum~\cite{luscher86:finiteb,luscher90:finite,luscher91:finitea,luscher91:finiteb}.

 Owing to the advent of L\"uscher's formula, various lattice studies, both quenched~\cite{Gupta:1993rn,Fukugita:1994ve,Aoki:1999pt,Aoki:2002in,Liu:2001ss,Du:2004ib,Aoki:2005uf,Aoki:2002ny} and unquenched~\cite{Yamazaki:2004qb,Beane:2005rj,PhysRevD.77.094507,PhysRevD.81.074506,Feng2010268,PhysRevD.86.034031,PhysRevD.87.034505,PhysRevD.87.054502,PhysRevLett.111.222001,PhysRevLett.111.192001}, have been performed over the years to investigate the scattering of hadrons.
The original L\"uscher's formula was derived for systems of two identical spinless particles in center-of-mass (COM) frame with periodic boundary condition in cubic box.
It restrains the
applicability of the formalism to general hadron scattering.
To overcome the difficulties, one can of course consider
 asymmetric volumes~\cite{Li:2003jn,Feng:2004ua}, or boosting the
 system to a frame that is different from COM~\cite{Rummukainen:1995vs,XuFeng:2011,Davoudi:2011md,ZiwenFu2012,Gockeler:2012yj}.
 Both will enhance the energy resolution of the problem. Another possible
 generalization is to use the so-called twisted boundary
 conditions advocated in
 Refs.~\cite{Bedaque:2004ax,Bedaque:2004kc,Sachrajda:2004mi,deDivitiis:2004kq}.
 All these generalizations enable one to compute the scattering phases in the low-momentum range which benefits for studying the property of low energy hadron-hadron scattering.
Then, generalizations to particles with spin~\cite{Beane:2006mx,Beane:2003da,Meng:2003gm} are also possible. For
 example, in Refs.~\cite{Bernard:2008ax,Ishizuka:2009bx}, L\"{u}scher's formulas
 have been extended to elastic scattering of baryons.

 In a cubic box, the three momenta of a single particle are quantized according to $\bk=(2\pi/L)\bn$
 with $\bn\in\mathcal{Z}^3$, where $\mathcal{Z}^3$ is the set of 3-tuples of integers.
 In real simulations, large values
of $L$ is needed to control lattice artifacts because of these non-zero momentum modes.
However, in the cubic lattice, many low
momentum modes are energy degenerate such as modes $(1, 0, 0)$, $(0, 1, 0)$ and $(0, 0, 1)$ since
they are related to each another by the cubic symmetry.
This means that the second
lowest energy level of the particle with non-vanishing momentum corresponds
to $\bn = (1, 1, 0)$. If one would like to measure these states on the lattice, even
larger values of $L$ should be used. One way to remedy this is to use an elongated box.
The technique of asymmetric box has first been applied to the calculation of pion-pion elastic scattering \cite{Li:2007ey} and later used in the study of $D^{\ast+}$-$D_1^0$ system \cite{Meng:2009qt}. In a recent
comprehensive study, Frank X. Lee and Andrei Alexandru have derived L\"uscher phase shift formulas for mesons and baryons in elongated boxes~\cite{Lee:2017igf}.
In their work, various scenarios, such as moving and zero-momentum states in cubic and elongated
boxes, are systematically studied and relations between them are also clarified. The derived formulas are applicable to a wide
set of meson-meson and meson-baryon elastic scattering processes. Therefore they can be applied to investigate various resonances, such as $a_1$, $\Delta$, Roper and so on.
In this paper, we would like to synthesize the above mentioned generalizations
 by trying to seek phase shift formulas that are applicable for baryon-baryon scattering in the elongated box which is a specific asymmetric box whose dimensions are
 $L\times{L}\times\eta{L}$, where $\eta$ is the elongation factor in the $z$-direction.

The organization of the paper is as follows. In Sec.~\ref{sec:baryon-baryon com},
 we start out by discussing scattering phase shift using a quantum-mechanical model in COM frame in an elongated box. Then, in Sec.~\ref{sec:generalizations},
 we generalize the scattering phase shift formulas in the elongated box: in subsection~\ref{sec:baryon-baryon mf}, we provide scattering phase shift formulas for baryon-baryon scattering in MF, and in subsection~\ref{sec:baryon-baryon tbc}, we show the scattering phase shift formulas for baryon-baryon scattering in COM frame with twisted boundary condition. In Sec.~\ref{sec:baryon-baryon cubic}, we obtain the results in a cubic box, and then perform the consistency checks and have confirmed the validation on the elongated results by comparing the two cases.
 In Sec.~\ref{sec:conclude}, we give a brief summary, and also discuss the possible
 applications of our derived formulas in real lattice simulations.

\section{Phase shift formulas for baryon-baryon scattering in COM frame in elongated boxes}
\label{sec:baryon-baryon com}
In this section, we try to obtain phase shift formulas that are applicable for baryon-baryon scattering in COM frame in elongated boxes.
Basically, we first derive the scattering phase shift formulas for baryon-baryon scattering in a cubic box in COM frame, and then we generalize the formulas
to the case in elongated boxes.
\subsection{Brief review of the basic L\"uscher's formula}
In nonrelativistic quantum mechanics,
after factoring out the center-of-mass motions, the asymptotic
form of the wave function is written as
\beq
\psi_{s\nu}(\mathbf{r})\overset{r\rightarrow \infty}{\longrightarrow}\left(
 \chi_{s\nu}e^{i\bk\cdot\br}
 +\sum_{s^{\prime}\nu^{\prime}}\chi_{s^{\prime}\nu^{\prime}}M_{s^{\prime}\nu^{\prime};s\nu}
\frac{e^{ikr}}{r}
 \right)\;.\label{wave_1}
 \eeq
In the remote past, the wave function reduces to an incident plane wave with prescribed quantum
numbers. Here $\chi_{s\nu}$ designates spin-wavefunction which is an eigenstate of spin angular
 momentum of $s^2$ and $\nu$ with the eigenvalues given by $s=0,\nu=0$ (singlet state),
 or $s=1,\nu=1,-1,0$ (triplet state).
 $M_{s^{\prime}\nu^{\prime},s\nu}$ is the scattering amplitude.
 When one chooses the z-axis to coincide with $\bk$, the scattering amplitudes introduced above is
 related to the $S$-matrix elements as~\cite{Newton}
 \beq
 M_{s^{\prime}\nu^{\prime};s\nu}(\mathbf{\hat{k}\cdot{\hat{r}}})&=&\frac{1}{2ik}
 \sum_{l^{\prime}=0}^\infty\sum_{l=0}^\infty\sum_{J=l-1}^{l+1}\sqrt{4\pi(2l+1)}i^{(l-l^{\prime})}\nonumber\\
 & &(S_{l^{\prime}s^{\prime};ls}^{J}-
 \delta_{l^{\prime}l}\delta_{s^{\prime}s})\langle{J}M|l^{\prime}m^{\prime};s^{\prime}\nu^{\prime}\rangle\nonumber\\
 & &\langle{J}M|l0;s\nu\rangle{Y_{l^{\prime}m^{\prime}}(\mathbf{\hat{r}})}
 \;.\label{NR_amplitude_1}
 \eeq
From the Clebsch-Gordan coefficients
 $\langle{J}M|l^{\prime}m^{\prime};s^{\prime}\nu^{\prime}\rangle$ and $\langle{J}M|l0;s\nu\rangle$, we find
 $M=\nu$ and $m=M-\nu^{\prime}$.
 In the previous equation, we have ignored the sum over  $M$ and $m$
 for a given pair of $\nu$ and $\nu^{\prime}$.
It is well-known that the phase shift enters via $S$-matrix.
L\"uscher obtained a relation, which relates the energy of two-particle state
 in a finite box of size $L$, i.e., $E(L)$, to the elastic scattering phase
 $\delta(E(L))$ of the two particles in the continuum. This inspires us that if one calculates two-particle scattering phase shifts,
 one should derive the corresponding L\"uscher's formula firstly.

With the help of Eq.~(\ref{wave_1}) and Eq.~(\ref{NR_amplitude_1}), we obtain the following asymptotic
form of the wave function as
 \beq
 \label{eq:wavefunction_intermsof_W}
 \psi_{s\nu}(\mathbf{r})=\sum_{s^{\prime}JMll^{\prime}}\sqrt{4\pi(2l+1)}W^{J}_{l^{\prime}s^{\prime};ls}
 \langle{J}M|l0;s\nu\rangle{Y^{l^{\prime}s^{\prime}}_{JM}(\hat{\mathbf{r}})}\;,
 \eeq
 where $Y^{l^{\prime}s^{\prime}}_{JM}(\hat{\mathbf{r}})$ is the spin spherical harmonics
 whose explicit form is given by
 \beq
 Y^{ls}_{JM}(\hat{\mathbf{r}})&=&\sum_{m\nu}Y_{lm}(\hat{\mathbf{r}})\chi_{s\nu}\langle{J}M|lm;s\nu\rangle.
 \eeq
 In Eq.~(\ref{eq:wavefunction_intermsof_W}), $W^{J}_{l^{\prime}s^{\prime};ls}(r)$
 is the radial wave function of the two-particle scattering state.
 In the large $r$ region, when the two-particle interaction is ignored and the potential vanishes, the wave function $W^{J}_{l^{\prime}s^{\prime};ls}(r)$ has the following asymptotic form:
 \beq
 W^{J}_{l^{\prime}s^{\prime};ls}(r)&=&
 \frac{1}{2ikr}i^{(l-l^{\prime})}
 [S^{J}_{l^{\prime}s^{\prime};ls}e^{ikr}+(-1)^{l+1}e^{-ikr}\delta_{ll^{\prime}}\delta_{ss^{\prime}}].\nonumber\\
 \;
 \eeq

 For baryon-baryon scattering, we enclose the two-particle system in a cubic box of size $L$ with periodic boundary condition.
 In the outer region where the potential vanishes, the wave function becomes
 \be
 \psi(\mathbf{r})=\sum_{s^{\prime}JMll^{\prime}s}
 \left[F_{JMls}W^{J}_{l^{\prime}s^{\prime};ls}(r)\right]
 Y_{JM}^{l^{\prime}s^{\prime}}(\mathbf{\hat{r}})\;.\label{wave_3}
 \ee
 Then, we can expand the wave function into a linear superposition of
 the singular periodic solutions, i.e., $G_{JMls}(\mathbf{r};k^{2})$,
 of the Helmholtz equation in the outer region. Thus, one can get
 \beq
 \psi(\mathbf{r})&=&
 \overset{1}{\underset{s=0}{\sum}}
 \overset{\infty}{\underset{l=0}{\sum}}\overset{l+1}
 {\underset{J=l-1}{\sum}}\overset{J}{\underset{M=-J}{\sum}}
 \upsilon_{JMls}G_{JMls}(\mathbf{r};k^{2})
\;. \label{wave_4}
 \eeq
 Here, $G_{JMls}(\mathbf{r};k^{2})$ can be further expanded in terms
 of spherical harmonics
 \beq
 G_{JMls}&=&\frac{(-1)^lk^{l+1}}{4\pi}(Y^{ls}_{JM}n_l(kr)\nonumber\\
 & &+\sum_{J^{\prime}M^{\prime}l^{\prime}}
 \mathcal{M}^{\mathbf{c}(s)}_{JMl;J^{\prime}M^{\prime}l^{\prime}}(k^2)
 Y^{l^{\prime}s}_{J^{\prime}M^{\prime}}j_{l^{\prime}}(kr))\;,\label{Green}
 \eeq
where the explicit form of $\mathcal{M}^{\mathbf{c}(s)}_{JMl;J^{\prime}M^{\prime}l^{\prime}}(q^2)$ with $\mathbf{k}=(2\pi/L)\mathbf{q}$ is written as
  \beq
 \mathcal{M}^{\mathbf{c}(s)}_{JMl;J^{\prime}M^{\prime}l^{\prime}}(q^2)&=&
 \sum_{mm'\nu}\langle{J}M|lm;s\nu\rangle\langle{J^{\prime}}M^{\prime}|l^{\prime}m^{\prime};s\nu\rangle\nonumber\\
 & &\times{\mathcal{M}}^{\mathbf{c}}_{lm;l^{\prime}m^{\prime}}(q^2),
 \label{M_b_c}
 \eeq
 with
 \beq
 \mathcal{M}_{lm;l^{\prime}m^{\prime}}^{\mathbf{c}}(q^2)&=&\sum_{t=|l-l^{\prime}|}^{l+l^{\prime}}\sum_{t_z=-t}^{t}\frac{(-1)^{l}i^{l+l^{\prime}}}{\pi^{3/2}{q^{t+1}}}\nonumber\\
 & &\times{Z}^{\mathbf{c}}_{tt_z}(q^2,\eta)\langle{l0t0}|{l^{\prime}0}\rangle\langle{lmtt_z}|{l^{\prime}m^{\prime}}\rangle\nonumber\\
 & &\times{\sqrt{\frac{(2l+1)(2t+1)}{(2l^{\prime}+1)}}}.
 \label{M_m_c}
 \eeq
For convenience, we introduce the short-hand function $\omega^{\mathbf{c}}_{lm}(q^2)$ for zeta function,
\beq
\omega^{\mathbf{c}}_{lm}(q^2)=\frac{Z^{\mathbf{c}}_{lm}(q^2)}{\pi^{3/2}{q}^{l+1}}.
\label{eq.Mw_c}
\eeq
Then the zeta function in cubic boxes is written as
\beq
Z^{\mathbf{c}}_{lm}(q^2)=\sum_{\mathbf{n}}\frac{\mathcal{Y}_{lm}(\mathbf{n})}{\mathbf{n}^2-q^2},
\eeq
where $\mathcal{Y}_{lm}(\mathbf{n})=\mathbf{n}^lY_{lm}(\theta,\phi)$ with $\mathbf{n}=(n_x,n_y,n_z)\in\mathcal{Z}^3$,
and $\mathcal{Z}^3$ is the set of 3-tuples of integers.
 One can get four linear equations of the coefficients by
 comparing Eq.~(\ref{wave_3}) with Eq.~(\ref{wave_4}).
 If there exist non-trivial solutions
 for them, the determinant of the corresponding matrix
 must vanish which leads to the basic form of L\"uscher's formula, i.e., Eq.~(\ref{eq:luscher_general_form}).
 \bw
 \beq
 \label{eq:luscher_general_form}
 \left|
 \sum_{l^{\prime\prime}}(S^{J}_{l^{\prime\prime}s^{\prime};ls}
 -\delta_{ll^{\prime\prime}}\delta_{ss^{\prime}})
 \mathcal{M}^{\mathbf{c}(s^{\prime})}_{JMl^{\prime\prime};J^{\prime}M^{\prime}l^{\prime}}(k^2)
 -i\delta_{JJ^{\prime}}\delta_{MM^{\prime}}(S^{J}_{l^{\prime}s^{\prime};ls}+\delta_{ll^{\prime}}\delta_{ss^{\prime}})
 \right|
 =0\;.
 \eeq
 \ew
 \subsection{Extension in elongated boxes in COM frame}
In the following, we generalize the basic form of L\"uscher's formula to that in an elongated box.
In this case, one can
expand the wave function of the system in the outer
region as series of the modified matrix, whose element is $\mathcal{M}^{(s)}_{JMl;J^{\prime}M^{\prime}l^{\prime}}(q^2,\eta)$, i.e.,
 \beq
 \mathcal{M}^{(s)}_{JMl;J^{\prime}M^{\prime}l^{\prime}}(q^2,\eta)&=&
 \sum_{mm'\nu}\langle{J}M|lm;s\nu\rangle\langle{J^{\prime}}M^{\prime}|l^{\prime}m^{\prime};s\nu\rangle\nonumber\\
 & &\times{\mathcal{M}}_{lm;l^{\prime}m^{\prime}}(q^2,\eta),
 \label{M_b}
 \eeq
 \beq
 \mathcal{M}_{lm;l^{\prime}m^{\prime}}(q^2,\eta)&=&\sum_{t=|l-l^{\prime}|}^{l+l^{\prime}}\sum_{t_z=-t}^{t}\frac{(-1)^{l}i^{l+l^{\prime}}}{\pi^{3/2}\eta{q^{t+1}}}\nonumber\\
 & &\times{Z}_{tt_z}(1,q^2,\eta)\langle{l0t0}|{l^{\prime}0}\rangle\langle{lmtt_z}|{l^{\prime}m^{\prime}}\rangle\nonumber\\
 & &\times{\sqrt{\frac{(2l+1)(2t+1)}{(2l^{\prime}+1)}}}.
 \label{M_m}
 \eeq
Hereafter for convenience, we introduce the short-hand function $\omega_{lm}(q^2,\eta)$ for zeta function as follows:
\beq
\omega_{lm}(q^2,\eta)=\frac{Z_{lm}(q^2,\eta)}{\pi^{3/2}\eta{q}^{l+1}}
\label{eq.Mw}
\eeq
Then the generalized zeta function for z-elongated boxes is given by
\beq
Z_{lm}(q^2,\eta)=\sum_{\mathbf{n}}\frac{\mathcal{Y}_{lm}(\mathbf{\tilde{n}})}{\mathbf{\tilde{n}}^2-q^2},
\eeq
where $\mathcal{Y}_{lm}(\mathbf{\tilde{n}})=\mathbf{\tilde{n}}^lY_{lm}(\theta,\phi)$. Here, the modified index $\mathbf{\tilde{n}}$ is $\mathbf{\tilde{n}}=(n_x,n_y,n_z/\eta)$.
Apart from the above-mentioned substitutions,
the extra attention should also be paid to the
difference in symmetry. In order to discuss L\"uscher's
formula in the elongated box, the symmetry of two-particle system is no longer $O_h$ but reduces to $D_{4h}$ group.
This group contains $16$ elements that can be divided into the following
ten conjugate classes: $A_1^{\pm}$,$A_2^{\pm}$,$B_1^{\pm}$,$B_2^{\pm}$, and $E^{\pm}$. For instance, for $J=0$, $1$,
 $2$ when the  cutoff momentum $\Lambda=2$, the decomposition into irreducible
 representation is given by $0^{\pm}=A_1^{\pm}$, $1^{\pm}=A_2^{\pm}\oplus{E^{\pm}}$,
 $2^{\pm}=A_1^{\pm}\oplus{B_1^{\pm}}\oplus{B_2^{\pm}}\oplus{E^{\pm}}$, respectively~\cite{Lee:2017igf}.
In a definite irreducible representation of the $D_{4h}$ group,
the basis vectors are labeled as $|\Gamma,\xi,J,l,s,n\rangle$, where $\Gamma$ denotes the representation;
$\xi$ runs from $1$ to the number of the dimensions, and $n$ runs from $1$ to the multiplicity of the representation.
This basis can be expressed by linear combinations of $|JMls\rangle$.
The corresponding matrix $\mathcal{M}$ is diagonal with respect to $\Gamma$
and $\xi$ by Schur's lemma~\cite{luscher91:finitea}.
 If there is no multiplicity, the labels $n=1$ and $n^{\prime}=1$ can be dropped.
 Therefore, in a definite symmetry sector $\Gamma$, the explicit form of L\"uscher's formula is given by Eq.~(\ref{eq:luscher_general_form2}).
\bw
 \be
  \label{eq:luscher_general_form2}
 \left|
 \sum_{l^{\prime\prime}}(S^{J}_{l^{\prime\prime}s^{\prime};ls}
 -\delta_{ll^{\prime\prime}}\delta_{ss^{\prime}})
 \mathcal{M}^{(s^{\prime})}_{Jl^{\prime\prime};J^{\prime}l^{\prime}}(\Gamma)
 -i\delta_{JJ^{\prime}}(S^{J}_{l^{\prime}s^{\prime};ls}+\delta_{ll^{\prime}}\delta_{ss^{\prime}})
 \right|
 =0\;.
 \ee
\ew

Before writing out the explicit form of the phase shift formula, one should exploit the symmetry properties to simplify the $\mathcal{M}$ matrix and the parity of the two-particle system.
On the one hand, the matrix is hermitian which constrains half of the off-diagonal elements. Furthermore, a lot of short-hand functions  $\omega_{lm}(q^2,\eta)$ vanish to satisfy certain constraints, which can be traced back to how the zeta function behaves under the symmetry operations in the elongated box. The properties follow Ref. \cite{Lee:2017igf}.
On the other hand, the total angular
momentum for two scattering particles with spin $s_1$ and $s_2$ is $J=s_1+s_2+l$ with $l$ the relative orbital angular momentum dubbed ``partial waves''. For the asymptotic states,
the two particles are far
away from each other so that they are not interacting. Then, we can label the states with $s$, $s_i$, and $s_z$, where $s=s_1+s_2$ is the total spin. The total angular momentum $J$ is conserved during the scattering process, but both $l$ and $s$ will change. For the baryon-baryon scattering considered in this paper, the total spin $s$ can take $0$ (singlet state) or $1$ (triplet states). Moreover, the orbital angular momentum also remains fixed for our cases: In the case with $s=0$ we have $l=J$ which is conserved, and the parity is simply $(-1)^J$.
When $s=1$ for a given $J$, $l$ takes three different values, i.e., $l=J+1,J-1,J$. The first two have the same parity $(-1)^{J+1}$, and the parity for third one is $(-1)^J$.
The total parity of the two-particle
state is equal to $P_{tol}=P_1P_2(-1)^l$, where $P_1$ and $P_2$ are
the intrinsic parities of the two scattering particles.
For simplicity we assume that the intrinsic parity $P_1P_2$ is positive, then the total parity is $(-1)^l$.
For parity-conserving theories like QCD, there is no scattering between states with opposite parity.
Then we divide L\"uscher's formulas into the case A and case B corresponding to the states with parity $(-1)^{J+1}$ and parity $(-1)^{J}$ respectively.

\subsection{Application to case A}
 In this case ($s=s'=1$, $l=J\pm1$), L\"uscher's formula becomes
 \be
 \left|
 \sum_{l^{\prime\prime}}(S^{J}_{l^{\prime\prime}1;l1}
 -\delta_{ll^{\prime\prime}})
 \mathcal{M}^{(1)}_{Jl^{\prime\prime};J^{\prime}l^{\prime}}
 -i\delta_{JJ^{\prime}}(S^{J}_{l^{\prime}1;l1}+\delta_{ll^{\prime}})
 \right|
 =0\;.\label{casea}
 \ee

According to non-zero matrix elements $\omega_{lm}$ are given in Tab.~\ref{table_symmetry1}, one can obtain phase shift in the definite symmetry.
Then we list the phase shift with different irreducible representations of the $D_{4h}$ group.

\begin{table*}[!htb]
\begin{center}
\caption{$D_{4h}$ symmetry group for angular momentum up to $J=2$ and $l=2$. The results are separated according to parity $(-1)^{J\pm1}$}
\label{table_symmetry1}       
\begin{tabular}{llll}
\hline\hline
$\Gamma$ &$Jl$  &$J^{\prime}l^{\prime}$ &$\mathcal{M}^{(1)}_{Jl;J^{\prime}l^{\prime}}(\Gamma)$\\
\noalign{\smallskip}\hline\noalign{\smallskip}
$A_1^{-}$ &$01$  &$01$
          &$\omega_{00}-\frac{2\sqrt{30}}{15}\omega_{22}$\\
          &$01$  &$21$
          &$-\frac{\sqrt{10}}{5}\omega_{20}-\frac{2\sqrt{15}}{15}\omega_{22}$\\
          &$21$  &$21$
          &$\omega_{00}+\frac{\sqrt{5}}{5}\omega_{20}-\frac{\sqrt{30}}{15}\omega_{22}$\\
\hline
$A_2^{+}$ &$10$  &$10$
          &$\omega_{00}$\\
          &$10$  &$12$
          &$\frac{\sqrt{10}}{5}\omega_{20}$\\
          &$12$  &$12$
          &$\omega_{00}+\frac{\sqrt{5}}{5}\omega_{20}-\frac{3\sqrt{30}}{35}\omega_{22}-\frac{6\sqrt{10}}{35}\omega_{42}$\\
\hline
$B_1^{-}$ &$21$  &$21$
          &$\omega_{00}-\frac{\sqrt{5}}{5}\omega_{20}-\frac{\sqrt{30}}{5}\omega_{22}$\\
\hline
$B_2^{-}$ &$21$  &$21$
          &$\omega_{00}-\frac{\sqrt{5}}{5}\omega_{20}+\frac{\sqrt{30}}{5}\omega_{22}$\\
\hline
$E^{+}$   &$10$  &$10$
          &$\omega_{00}$\\
          &$10$  &$12$
          &$-\frac{\sqrt{10}}{10}\omega_{20}-\frac{\sqrt{15}}{5}\omega_{22}$\\
          &$12$  &$12$
          &$\omega_{00}-\frac{\sqrt{5}}{10}\omega_{20}-\frac{2\sqrt{30}}{35}\omega_{22}+\frac{3\sqrt{10}}{35}\omega_{42}$\\
\hline
$E^{-}$   &$21$  &$21$
          &$\omega_{00}+\frac{\sqrt{5}}{10}\omega_{20}$\\
\hline\hline
\end{tabular}
\end{center}
\end{table*}

If we consider the explicit parity and also suppose that the cutoff angular momentum is $\Lambda=2$,
 the decomposition in this case becomes
 $0^{-}=A_1^{-}$, $1^{+}=A_2^{+}\oplus{E^{+}}$,
 $2^{-}=A_1^{-}\oplus{B_1^{-}}\oplus{B_2^{-}}\oplus{E^{-}}$.
For instance, in the $A_1^-$ representation, the phase shift formula is Eq.~(\ref{eq:A1-casea}).
\bw
\beq
\left|
\ba
 {cc}
\cot\delta_{01}-(\omega_{00}-\frac{2\sqrt{30}}{15}\omega_{22})   &-\frac{\sqrt{10}}{5}\omega_{20}-\frac{2\sqrt{15}}{15}\omega_{22}\\
-\frac{\sqrt{10}}{5}\omega_{20}-\frac{2\sqrt{15}}{15}\omega_{22}     &\cot\delta_{21}-(\omega_{00}+\frac{\sqrt{5}}{5}\omega_{20}-\frac{\sqrt{30}}{15}\omega_{22})
\ea
\right|=0
\label{eq:A1-casea}
\eeq
\ew
If we ignore the mixing with $J=2$, one can extract phase shift $\delta_{01}$ from
\be
\cot\delta_{01}=\omega_{00}-\frac{2\sqrt{30}}{15}\omega_{22}.
\ee
Next, if we want to calculate phaseshift $\delta_{21}$, we can consider representation $B_1^-$, $B_2^-$, and $E^-$.
The phase shift formula in these three representations are written as
\be
 \left\{ \begin{aligned}
\cot\delta_{21}=\omega_{00}-\frac{\sqrt{5}}{5}\omega_{20}-\frac{\sqrt{30}}{5}\omega_{22}\\
\cot\delta_{21}=\omega_{00}-\frac{\sqrt{5}}{5}\omega_{20}+\frac{\sqrt{30}}{5}\omega_{22}\\
\cot\delta_{21}=\omega_{00}+\frac{\sqrt{5}}{10}\omega_{20}\\
 \end{aligned} \right.\;.\nonumber\\
\ee
Finally, let us discuss the quantum number $J^P$ of the two-particle state which is $1^+$  in $A_2^+$ and $E^+$ representations, i.e., the total angular momentum $J$ is fixed for $1$.
There is a mixing with $l=J-1=0$ (S-wave) and $l=J+1=2$ (D-wave).
To proceed, we need to parameterize $S$-matrix.
First we note that
$S$ must be unitary (conservation of probability) and
symmetric (reciprocity) due to the
T-invariance of the strong interactions. Thus the $2\times2$
$S$-matrix is determined by three real parameters. We use the
``eigenphase convention'' of Blatt and Biedenharn~\cite{Blatt:1952zza} in Eq.~(\ref{eq.s_matrix}),
\begin{figure*}[!htb]
\be
S_{2\times2}=\left(
\ba
 {cc}
\cos\epsilon   &-\sin\epsilon\\
\sin\epsilon     &\cos\epsilon
\ea
\right)
\left(
\ba
{cc}
e^{2i\delta_\alpha}  &0\\
0         &e^{2i\delta_\beta}
\ea
\right)
\left(
 \ba
 {cc}
\cos\epsilon   &\sin\epsilon\\
-\sin\epsilon     &\cos\epsilon
\ea
\right)
\label{eq.s_matrix}
\ee
\end{figure*}
where $\delta_\alpha$ and $\delta_\beta$ are the scattering phase shifts corresponding to two eigenstates of the S-matrix called ``$\alpha$'' and ``$\beta$'' waves respectively. At low energies, the $\alpha$-wave is predominantly S-wave with a small admixture of the D-wave, while the
$\beta$-wave is predominantly D-wave with a small admixture of the S-wave.
Therefore, it hardly needs to be emphasized that the eigenphaseshifts
$\delta_\alpha$ and $\delta_\beta$ are not to be thought of as phase shifts for the states $l=0$ and $l=2$, respectively.
There are no such phase shifts due to that neither of these
two states is an eigenstate of the scattering matrix. The
``mixture parameter'' $\epsilon$ which determines the ``correct''
mixtures of these two states is an essential parameter in
the scattering matrix and enters explicitly into the
differential cross section.
Thus, in $A_2^+$ representation, the phase shift formula is shown in Eq.~(\ref{eq:A2+casea}),
\bw
\be
\left|
\left(
S_{2\times2}-I_{2\times2}
\right)
\left(
\ba
{cc}
\omega_{00}  &\frac{\sqrt{10}}{5}\omega_{20}\\
\frac{\sqrt{10}}{5}\omega_{20}         &\omega_{00}+\frac{\sqrt{5}}{5}\omega_{20}-\frac{3\sqrt{30}}{35}\omega_{22}-\frac{6\sqrt{10}}{35}\omega_{42}
\ea
\right)
 -i\left(
 S_{2\times2}+I_{2\times2}
\right)
\right|
 =0,
 \label{eq:A2+casea}
\ee
\ew
where $I_{2\times{2}}$ is a $2\times{2}$ unit matrix.
In $E^+$ representation, we should substitute $\mathcal{M}^{(1)}_{Jl^{\prime\prime};J^{\prime}l^{\prime}}$ according to Tab.~\ref{table_symmetry1} with Eq.~(\ref{eq:A2+casea}).

\subsection{Application to case B}
 For this case ($l=l^{\prime}=l^{\prime\prime}=J$, $s=0,1$), the L\"uscher's formula can be expressed as shown in Eq.~(\ref{Luscher_formula5}).
 \bw
 \be
 \left|
 \sum_{l^{\prime\prime}}(S^{J}_{l^{\prime\prime}s^{\prime};ls}
 -\delta_{ll^{\prime\prime}}\delta_{ss^{\prime}})
 \mathcal{M}^{(s^{\prime})}_{Jl^{\prime\prime};J^{\prime}l^{\prime}}\\
 -i\delta_{JJ^{\prime}}(S^{J}_{ls^{\prime};ls}+\delta_{ss^{\prime}})
 \right|
 =0\;.\label{Luscher_formula5}
 \ee
\ew
 If we consider the explicit parity and also suppose that the cutoff angular momentum is $\Lambda=2$,
 the decomposition in this case becomes
 $0^{+}=A_1^{+}$, $1^{-}=A_2^{-}\oplus{E^{-}}$,
 $2^{+}=A_1^{+}\oplus{B_1^{+}}\oplus{B_2^{+}}\oplus{E^{+}}$.
Then, we parameterize the  $S$-matrix again using the
"eigenphase convention" of Blatt and Biedenharn~\cite{Blatt:1952zza}, whose explicit form is similar to Eq.~(\ref{eq.s_matrix}).
Thus, $\delta_\alpha$ and $\delta_\beta$ are the scattering phase shifts corresponding to two eigenstates of the S-matrix called ¡°$\alpha$¡± and ¡°$\beta$¡± waves respectively. At low energies.
However, the $\alpha$-wave is predominantly $s=0$ with a small admixture of the $s=1$, while the
$\beta$-wave is predominantly $s=1$ with a small admixture of the $s=0$.
According to the non-zero matrix elements, one can obtain phase shift formula in a definite symmetry.
For example, if we focus on the $A^{+}_1$ representation with positive parity that corresponds $J=0$ and ignoring the index $l$ and $l^{\prime}$, both of which are unity,
the phase shift formula is Eq.~(\ref{eq:pfA1+caseb1}).
\bw
\be
\left|
\left(
S_{2\times2}-I_{2\times2}
\right)
\left(
\ba
{cc}
\omega_{00}+\frac{6}{7}\omega_{40}+\frac{2\sqrt{5}}{7}\omega_{20}   &0\\
0        &\omega_{00}-\frac{4}{7}\omega_{40}+\frac{\sqrt{5}}{7}\omega_{20}+\frac{\sqrt{30}}{7}\omega_{22}+\frac{2\sqrt{10}}{7}\omega_{42}
\ea
\right)
 -i\left(
 S_{2\times2}+I_{2\times2}
\right)
\right|
 =0
 \label{eq:pfA1+caseb1}
\ee
\ew
In addition, phase shifts in $B_1^+$, $B_2^+$, and $E^+$ are similar to Eq.~(\ref{eq:pfA1+caseb1}) expect that the matrix $\mathcal{M}^{(s^{\prime})}_{Jl^{\prime\prime};J^{\prime}l^{\prime}}$ in the equation are replaced by the corresponding one according to Tab.~\ref{table_symmetry2} and Tab.~\ref{table_symmetry3}.
Next, taking $A_2^-$ representation with negative parity as an example, the formula is written as Eq.~(\ref{eq:pfA2-caseb}).
\bw
\be
\left|
\left(
S_{2\times2}-I_{2\times2}
\right)
\left(
\ba
{cc}
\omega_{00}+\frac{2\sqrt{5}}{5}\omega_{20}   &0\\
0       &\omega_{00}-\frac{\sqrt{5}}{5}\omega_{20}+\frac{\sqrt{30}}{5}\omega_{22}
\ea
\right)
 -i\left(
 S_{2\times2}+I_{2\times2}
\right)
\right|
=0
\label{eq:pfA2-caseb}
\ee
\ew
The formula in $E^-$ representation is similar to Eq.~(\ref{eq:pfA2-caseb}) expect that the matrix $\mathcal{M}^{(s^{\prime})}_{Jl^{\prime\prime};J^{\prime}l^{\prime}}$ in the equation is replaced by the corresponding one according to Tab.~\ref{table_symmetry2} and Tab.~\ref{table_symmetry3}.

In the above discussion, we have not taken into account the possibility
for the identical nature of the two scattering particles, and thus the
singlet-triplet transition within the same parity is allowed.
However, for the two identical particles, the singlet-triplet transition is forbidden since the singlet state has
an antisymmetric spin wave function which then requires a symmetric spatial one
that necessarily has positive parity while the triplet state has the opposite parity.
Below, we list L\"uscher formulas for $s=s^{\prime}=0$ and $s=s^{\prime}=1$ cases respectively:
 \be
 \left|
 \sum_{l^{\prime\prime}}(S^{J}_{l^{\prime\prime}0;l0}
 -\delta_{ll^{\prime\prime}})
 \mathcal{M}^{(0)}_{Jl^{\prime\prime};J^{\prime}l^{\prime}}\\
 -i\delta_{JJ^{\prime}}(S^{J}_{l0;l0}+1)
 \right|
 =0\;.\label{Lusher_formulae_1}
 \ee
 \be
 \left|
 \sum_{l^{\prime\prime}}(S^{J}_{l^{\prime\prime}1;l1}
 -\delta_{ll^{\prime\prime}})
 \mathcal{M}^{(1)}_{Jl^{\prime\prime};J^{\prime}l^{\prime}}\\
 -i\delta_{JJ^{\prime}}(S^{J}_{l1;l1}+1)
 \right|
 =0\;.\label{Lusher_formulae_2}
 \ee
  From these explicit expressions, we find they are quite similar to those in the
 case of meson-meson scattering. In particular, L\"uscher's formula for $s=s^{\prime}=0$ is same with the meson-meson scattering case. Then, we only discuss the case for $s=s^{\prime}=1$.
 In this case, we denote phase shift as $\delta_{J}$ (we have ignored the index $l$ due to $J=l$ in this case) in the definite representation.
 According to the non-zero matrix elements listed in Tab.~\ref{table_symmetry2}, one can obtain phase shift $\delta_{J}$ in the definite representation.
\begin{table*}[!htb]
\begin{center}
\caption{$D_{4h}$ symmetry group for angular momentum up to $J=2$ and $l=2$. The results are separated according to parity $(-1)^{J}$.$s=s^{\prime}=1$}
\label{table_symmetry2}       
\begin{tabular}{llll}
\hline\hline
$\Gamma$ &$Jl$  &$J^{\prime}l^{\prime}$ &$\mathcal{M}^{(1)}_{Jl;J^{\prime}l^{\prime}}(\Gamma)$\\
\noalign{\smallskip}\hline\noalign{\smallskip}
$A_1^{+}$ &$22$  &$22$
          &$\omega_{00}-\frac{4}{7}\omega_{40}+\frac{\sqrt{5}}{7}\omega_{20}+\frac{\sqrt{30}}{7}\omega_{22}+\frac{2\sqrt{10}}{7}\omega_{42}$\\
\hline
$A_2^{-}$ &$11$  &$11$
          &$\omega_{00}-\frac{\sqrt{5}}{5}\omega_{20}+\frac{\sqrt{30}}{5}\omega_{22}$\\
\hline
$B_1^{+}$ &$22$  &$22$
          &$\omega_{00}-\frac{2}{21}\omega_{40}-\frac{\sqrt{5}}{7}\omega_{20}+\frac{\sqrt{30}}{21}\omega_{22}+\frac{2\sqrt{10}}{21}\omega_{42}-\frac{2\sqrt{70}}{21}\omega_{44}$\\
\hline
$B_2^{+}$ &$22$  &$22$
          &$\omega_{00}-\frac{2}{21}\omega_{40}-\frac{\sqrt{5}}{7}\omega_{20}-\frac{\sqrt{30}}{21}\omega_{22}-\frac{2\sqrt{10}}{21}\omega_{42}+\frac{2\sqrt{70}}{21}\omega_{44}$\\
\hline
$E^{-}$   &$11$  &$11$
          &$\omega_{00}+\frac{\sqrt{5}}{10}\omega_{20}$\\
\hline
$E^{+}$   &$22$  &$22$
          &$\omega_{00}+\frac{8}{21}\omega_{40}+\frac{\sqrt{5}}{14}\omega_{20}+\frac{2\sqrt{30}}{21}\omega_{22}-\frac{\sqrt{10}}{7}\omega_{42}$\\
\hline\hline
\end{tabular}
\end{center}
\end{table*}
\begin{table*}[!htb]
\begin{center}
\caption{$D_{4h}$ symmetry group for angular momentum up to $J=2$ and $l=2$. The results are separated according to parity $(-1)^{J}$.$s=s^{\prime}=0$}
\label{table_symmetry3}       
\begin{tabular}{llll}
\hline\hline
$\Gamma$ &$Jl$  &$J^{\prime}l^{\prime}$ &$\mathcal{M}^{(0)}_{Jl;J^{\prime}l^{\prime}}(\Gamma)$\\
\noalign{\smallskip}\hline\noalign{\smallskip}
$A_1^{+}$ &$22$  &$22$    &$\omega_{00}+\frac{6}{7}\omega_{40}+\frac{2\sqrt{5}}{7}\omega_{20}$\\
\hline
$A_2^{-}$ &$11$  &$11$    &$\omega_{00}+\frac{2\sqrt{5}}{5}\omega_{20}$\\
\hline
$B_1^{+}$ &$22$  &$22$
          &$\omega_{00}+\frac{1}{7}\omega_{40}-\frac{2\sqrt{5}}{7}\omega_{20}+\frac{\sqrt{70}}{7}\omega_{44}$\\
\hline
$B_2^{+}$ &$22$  &$22$
          &$\omega_{00}+\frac{1}{7}\omega_{40}-\frac{2\sqrt{5}}{7}\omega_{20}-\frac{\sqrt{70}}{7}\omega_{44}$\\
\hline
$E^{-}$   &$11$  &$11$
          &$\omega_{00}-\frac{\sqrt{5}}{5}\omega_{20}$\\
\hline
$E^{+}$   &$22$  &$22$
          &$\omega_{00}+\frac{\sqrt{5}}{7}\omega_{20}-\frac{4}{7}\omega_{40}$\\
\hline\hline
\end{tabular}
\end{center}
\end{table*}

If we focus on phase shift $\delta_2$, one can consider $A_1^+$, $B_1^+$, $B_2^+$, and $E^+$ representations. The phase shift formulas in these representations are similar with each other
expect that the matrix $\mathcal{M}^{\Gamma}_{Jl;J^{\prime}l^{\prime}}$ is different. We list the explicit form of these matrix in Tab.~{\ref{table_symmetry2}}.
For instance, we take $A_1^+$ representation as an example, where the phase shift formula is
\be
\cot\delta_2=\omega_{00}-\frac{4}{7}\omega_{40}+\frac{\sqrt{5}}{7}\omega_{20}+\frac{\sqrt{30}}{7}\omega_{22}+\frac{2\sqrt{10}}{7}\omega_{42}.
\ee
Then, if one wants to obtain $\delta_1$, one should consider $A_2^-$ and $E^-$ representations.
Here, we take $A_2^-$ representation as an example. In this case, the phase shift $\delta_1$ is contained in the flowing equation:
\be
\cot\delta_1=\omega_{00}-\frac{\sqrt{5}}{5}\omega_{20}+\frac{\sqrt{30}}{5}\omega_{22}.
\ee

\section{Some generalizations of phase shift formulas for baryon-baryon scattering in elongated boxes}
\label{sec:generalizations}
\subsection{Phase shift formulas for baryon-baryon scattering in MF}
\label{sec:baryon-baryon mf}
In this subsection, we extend the phase shift formula that has been obtained in the previous subsection for baryon-baryon
scattering to moving frames (MF) in an elongated box.
We will follow the notations in Ref.~\cite{Rummukainen:1995vs} below. We denote the four momenta of the two particles with periodic boundary conditions
in the lab frame by
 \be
 k=(E_1,\bk)\;,
 \;\;
 P-k=(E_2,\bP-\bk)\;,
 \ee
 where $E_1=\sqrt{\bk^2+m^{2}_1}$ and $E_2=\sqrt{(\bP-\bk)^2+m^{2}_2}$ are energies of the two particles respectively,
 $m_{1}$ and $m_{2}$ are the mass values of the
 two baryons respectively.
 The total three momentum $\mathbf{P}\neq 0$ of the two-particle system is
 quantized by the condition $\mathbf{P}=(2\pi/L)\mathbf{d}$
 with $\mathbf{d}\in \mathbbm{Z}^{3}$.
 The COM frame is then moving relative to the lab frame with a velocity
 \be
 \bv=\bP/(E_1+E_2)\;.
 \ee
 Then, we denote the momenta of the two particles with $\bk^{\ast}$ and $(-\bk^{\ast})$ for the two scattering particles.
 Thus, $\bk^{\ast}$ is related to $\bk$  by conventional Lorentz boost:
 \be
 \bk^{\ast\parallel}=\gamma(\bk^\parallel-\bv E_1)\;,
 \bk^{\ast\perp}=\bk^\perp\;,
 \ee
 where the symbol $\perp$ and $\parallel$ designate
 the components of the corresponding vector perpendicular
 and parallel to $\bv$, respectively.
 For simplicity, the above relation is also denoted
 by the shorthand notation: $\bk^{\ast}=\vec{\gamma}\bk$.
 A similar transformation relation holds for the other particles.

Following similar steps listed in Sec.~\ref{sec:baryon-baryon com}, we obtained the L\"uscher's formula in this case.
 L\"uscher's formula takes exactly the same form as Eq.~(\ref{eq:luscher_general_form})
 except that all the labels for $\mathcal{M}^{\mathbf{c}(s)}_{JMl;J^{\prime}M^{\prime}l^{\prime}}$ are replaced
 with the modified matrix elements $\mathcal{M}^{\mathbf{d}(s)}_{JMl;J^{\prime}M^{\prime}l^{\prime}}$.
 The explicit form for the matrix element is written as
 \beq
 \mathcal{M}^{\mathbf{d}(s)}_{JMl;J^{\prime}M^{\prime}l^{\prime}}(\boldsymbol{\kappa}^2,\eta)&=&\sum_{mm'\nu}\langle{J}M|lm;s\nu\rangle\langle{J^{\prime}}M^{\prime}|l^{\prime}m^{\prime};s\nu\rangle
 \nonumber\\
 & &\times{\mathcal{M}}^{\mathbf{d}}_{lm;l^{\prime}m^{\prime}}(\boldsymbol{\kappa}^2,\eta)
 \label{M_mf}
 \eeq
 with $\boldsymbol{\kappa}=\mathbf{k}L/(2\pi)$.
  The reduction matrix in moving frames has been obtained in
 Ref.~\cite{Rummukainen:1995vs,ZiwenFu2012}, whose explicit form is given by
 \beq
 {\mathcal{M}}^{\mathbf{d}}_{lm;l^{\prime}m^{\prime}}(\boldsymbol{\kappa}^2,\eta)&=&\sum_{t=|l-l^{\prime}|}^{l+l^{\prime}}\sum_{t^{\prime}=-t}^{t}
 \frac{(-1)^{l}i^{l+l^{\prime}}}{\gamma\pi^{3/2}\eta{\boldsymbol{\kappa}^{t+1}}}\nonumber\\
 & &\times{Z}^{\mathbf{d}}_{tt^{\prime}}(\boldsymbol{\kappa}^2,\eta)\langle{l0t0}|{l^{\prime}0}\rangle\langle{lmtt^{\prime}}|{l^{\prime}m^{\prime}}\rangle\nonumber\\
 & &\times\sqrt{\frac{(2l+1)(2t+1)}{(2l^{\prime}+1)}},
 \label{M_m_fm}
 \eeq
where $\gamma$ is the Lorentz factor of the boost, and the
 modified zeta function $Z^{\mathbf{d}}_{tt^{\prime}}(\boldsymbol{\kappa}^2,\eta)$
 is defined via
 \beq
Z^{\mathbf{d}}_{tt^{\prime}}(\boldsymbol{\kappa}^2,\eta)=\sum_{\mathbf{\tilde{n}}\in{\mathcal{P}_{\mathbf{d}}}}\frac{\mathcal{Y}_{tt^{\prime}}(\mathbf{\tilde{n}})}{\mathbf{\tilde{n}}^2-\boldsymbol{\kappa}^2}.
\eeq
In the above formulae, $\mathcal{P}_{\mathbf{d}}$ is the following
 set,
 \be
 \mathcal{P}_{\mathbf{d}}=\left\{\mathbf{\tilde{n}}\in\mathbbm{R}^3|\mathbf{\tilde{n}}
 =\gamma(\mathbf{n}+\frac{1}{2}\alpha{\mathbf{d}})
 \;\;\; \mathbf{n}\in{\mathbbm{Z}^{3}} \right\},
 \ee
 where $\mathbf{d}=(L/2\pi)\mathbf{P}$,
 $\alpha=1+\frac{w_{1\mathbf{k}}^{\ast2}-w_{2\mathbf{k}}^{\ast2}}{E^{\ast2}}$,
 and $w_{i\mathbf{k}}^{\ast2}=m^2_i+\mathbf{k}^{\ast2}$ with $i=1,2$ being the energy
 for the two scattering particles. We will also use $E^{\ast}$ to denote the
 total energy in the center of mass frame and introduce the short-hand function for zeta function
\beq
\omega^{\mathbf{d}}_{lm}(\boldsymbol{\kappa}^2,\eta)=\frac{Z^{\mathbf{d}}_{lm}(\boldsymbol{\kappa}^2,\eta)}{\gamma\pi^{3/2}\eta{\boldsymbol{\kappa}}^{l+1}}.
\eeq
In addition, an extra attention should also be paid to the difference in symmetries.
 In order to discuss phase shift formula in MF, we should introduce the space group $\mathcal{G}$,
 which is a semi-direct product of lattice translational group $\mathcal{T}$ and the cubic group $O$.
 The representations are characterized by the little group $\Gamma$
 and the corresponding total momentum $\mathbf{P}$.
 For example, for the case with $\mathbf{P}=\frac{2\pi}{L}\mathbf{e}_3$, the corresponding little group is $C_{4v}$~\cite{Gockeler:2012yj}.
 Then, following similar steps as in the COM frame, one can readily obtain the explicit
 formulas for the little group $C_{4v}$.
Before giving  the explicit form of phase shift, we should calculate non-vanished matrix of $\mathcal{M}^{\Gamma}_{Jl;J^{\prime}l^{\prime}}$ with the $C_{4v}$ group that is listed in Tab.~\ref{tab:C4v symmetry1} and Tab.~\ref{tab:C4v symmetry2}.
Then, following similar steps as in the Sec.~\ref{sec:baryon-baryon com}, one can
readily obtain the phase shift formulas in the moving frame. Here, we take phase shift formula in Case A($s=s'=1$, $l=J\pm1$) as an example.
For instance, in the $A_1$ representation, the phase shift formula is given by Eq.~(\ref{eq:A1caseamf}) if we ignore the mixing with $l=2$.
\bw
\beq
\left|
\ba
 {ccc}
\cot\delta_{01}-(\omega^{\mathbf{d}}_{00}-\frac{2\sqrt{30}}{15}\omega^{\mathbf{d}}_{22})   &\frac{2\sqrt{3}i}{3}\omega^{\mathbf{d}}_{11}+\frac{\sqrt{3}i}{3}\omega^{\mathbf{d}}_{10} &-\frac{\sqrt{10}}{5}\omega^{\mathbf{d}}_{20}-\frac{2\sqrt{15}}{15}\omega^{\mathbf{d}}_{22}\\
-\frac{2\sqrt{3}i}{3}\omega^{\mathbf{d}}_{11}-\frac{\sqrt{3}i}{3}\omega^{\mathbf{d}}_{10}     &\cot\delta_{10}-\omega^{\mathbf{d}}_{00} &\frac{\sqrt{6}i}{3}\omega^{\mathbf{d}}_{11}+\frac{\sqrt{6}i}{3}\omega^{\mathbf{d}}_{10}\\
-\frac{\sqrt{10}}{5}\omega^{\mathbf{d}}_{20}-\frac{2\sqrt{15}}{15}\omega^{\mathbf{d}}_{22} &-\frac{\sqrt{6}i}{3}\omega^{\mathbf{d}}_{11}-\frac{\sqrt{6}i}{3}\omega^{\mathbf{d}}_{10}
&\cot\delta_{21}-(\omega^{\mathbf{d}}_{00}+\frac{\sqrt{5}}{5}\omega^{\mathbf{d}}_{20}-\frac{\sqrt{30}}{15}\omega^{\mathbf{d}}_{22})
\ea
\right|=0
\label{eq:A1caseamf}
\eeq
\ew
We note that there is mixing between odd and even $J$ due to lack of parity in the boosted two-particle state.

\begin{table*}[!htb]
\begin{center}
\caption{$C_{4v}$ symmetry group for angular momentum up to $J=2$ and $l=2$. The results are separated according to parity $(-1)^{J\pm1}$.$s=s^{\prime}=1$}
\label{tab:C4v symmetry1}       
\begin{tabular}{llll}
\hline\hline
$\Gamma$ &$Jl$  &$J^{\prime}l^{\prime}$ &$\mathcal{M}^{\mathbf{d}(1)}_{Jl;J^{\prime}l^{\prime}}(\Gamma)$\\
\noalign{\smallskip}\hline\noalign{\smallskip}
$A_1$     &$01$  &$01$    &$\omega^{\mathbf{d}}_{00}-\frac{2\sqrt{30}}{15}\omega^{\mathbf{d}}_{22}$\\
          &$01$  &$10$    &$\frac{2\sqrt{3}i}{3}\omega^{\mathbf{d}}_{11}+\frac{\sqrt{3}i}{3}\omega^{\mathbf{d}}_{10}$\\
          &$01$  &$12$    &$\frac{\sqrt{150}i}{15}\omega^{\mathbf{d}}_{10}-\frac{\sqrt{150}i}{15}\omega^{\mathbf{d}}_{11}-\frac{2\sqrt{105}i}{35}\omega^{\mathbf{d}}_{32}$\\
          &$01$  &$21$    &$-\frac{\sqrt{10}}{5}\omega^{\mathbf{d}}_{20}-\frac{2\sqrt{15}}{15}\omega^{\mathbf{d}}_{22}$\\
          &$10$  &$10$    &$\omega^{\mathbf{d}}_{00}$\\
          &$10$  &$12$    &$\frac{\sqrt{10}}{5}\omega^{\mathbf{d}}_{20}$\\
          &$10$  &$21$    &$\frac{\sqrt{6}i}{3}\omega^{\mathbf{d}}_{11}+\frac{\sqrt{6}i}{3}\omega^{\mathbf{d}}_{10}$\\
          &$12$  &$12$    &$\omega^{\mathbf{d}}_{00}+\frac{\sqrt{5}}{5}\omega^{\mathbf{d}}_{20}-\frac{3\sqrt{30}}{35}\omega^{\mathbf{d}}_{22}-\frac{6\sqrt{10}}{35}\omega^{\mathbf{d}}_{42}$\\
          &$12$  &$21$    &$\frac{4\sqrt{3}i}{15}\omega^{\mathbf{d}}_{11}+\frac{\sqrt{3}i}{15}\omega^{\mathbf{d}}_{10}+\frac{9\sqrt{7}i}{35}\omega^{\mathbf{d}}_{30}+\frac{6\sqrt{42}i}{35}\omega^{\mathbf{d}}_{31}+\frac{\sqrt{210}i}{35}\omega^{\mathbf{d}}_{32}$\\
          &$21$  &$21$    &$\omega^{\mathbf{d}}_{00}+\frac{\sqrt{5}}{5}\omega^{\mathbf{d}}_{20}-\frac{\sqrt{30}}{15}\omega^{\mathbf{d}}_{22}$\\
\hline
$B_1$     &$21$  &$21$
          &$\omega^{\mathbf{d}}_{00}-\frac{\sqrt{5}}{5}\omega^{\mathbf{d}}_{20}-\frac{\sqrt{30}}{5}\omega^{\mathbf{d}}_{22}$\\
\hline
$B_2$     &$21$  &$21$
          &$\omega^{\mathbf{d}}_{00}-\frac{\sqrt{5}}{5}\omega^{\mathbf{d}}_{20}+\frac{\sqrt{30}}{5}\omega^{\mathbf{d}}_{22}$\\
\hline
$E$       &$10$  &$10$   &$\omega^{\mathbf{d}}_{00}$\\
          &$10$  &$12$   &$-\frac{\sqrt{10}}{10}\omega^{\mathbf{d}}_{20}-\frac{\sqrt{15}}{5}\omega^{\mathbf{d}}_{22}$\\
          &$10$  &$21$   &$\frac{\sqrt{2}i}{2}\omega^{\mathbf{d}}_{10}+\frac{\sqrt{2}i}{2}\omega^{\mathbf{d}}_{11}$\\
          &$12$  &$12$   &$\omega^{\mathbf{d}}_{00}-\frac{\sqrt{5}}{10}\omega^{\mathbf{d}}_{20}-\frac{2\sqrt{30}}{35}\omega^{\mathbf{d}}_{22}+\frac{3\sqrt{10}}{35}\omega^{\mathbf{d}}_{42}$\\
          &$12$  &$21$   &$\frac{2i}{5}\omega^{\mathbf{d}}_{11}+\frac{i}{10}\omega^{\mathbf{d}}_{10}-\frac{6\sqrt{14}i}{35}\omega^{\mathbf{d}}_{31}-\frac{3\sqrt{21}i}{35}\omega^{\mathbf{d}}_{30}-\frac{3\sqrt{70}i}{70}\omega^{\mathbf{d}}_{32}$\\
          &$21$  &$21$   &$\omega^{\mathbf{d}}_{00}+\frac{\sqrt{5}}{10}\omega^{\mathbf{d}}_{20}$\\
\hline\hline
\end{tabular}
\end{center}
\end{table*}

\begin{table*}[!htb]
\begin{center}
\caption{$C_{4v}$ symmetry group for angular momentum up to $J=2$ and $l=2$. The results are separated according to parity $(-1)^{J}$.$s=s^{\prime}=1$}
\label{tab:C4v symmetry2}       
\begin{tabular}{llll}
\hline\hline
$\Gamma$ &$Jl$  &$J^{\prime}l^{\prime}$ &$\mathcal{M}^{\mathbf{d}(1)}_{Jl;J^{\prime}l^{\prime}}(\Gamma)$\\
\noalign{\smallskip}\hline\noalign{\smallskip}
$A_1$     &$11$  &$11$    &$\omega^{\mathbf{d}}_{00}-\frac{\sqrt{5}}{5}\omega^{\mathbf{d}}_{20}+\frac{\sqrt{30}}{5}\omega^{\mathbf{d}}_{22}$\\
          &$11$  &$22$    &$\frac{\sqrt{15}i}{5}\omega^{\mathbf{d}}_{10}+\frac{3\sqrt{35}i}{35}\omega^{\mathbf{d}}_{30}-\frac{\sqrt{42}i}{7}\omega^{\mathbf{d}}_{32}$\\
          &$22$  &$22$    &$\omega^{\mathbf{d}}_{00}-\frac{4}{7}\omega^{\mathbf{d}}_{40}+\frac{2\sqrt{10}}{7}\omega^{\mathbf{d}}_{42}+\frac{\sqrt{30}}{7}\omega^{\mathbf{d}}_{22}+\frac{\sqrt{5}}{7}\omega^{\mathbf{d}}_{20}$\\
\hline
$B_1$     &$22$  &$22$
          &$\omega^{\mathbf{d}}_{00}-\frac{2}{21}\omega^{\mathbf{d}}_{40}-\frac{\sqrt{5}}{7}\omega^{\mathbf{d}}_{20}+\frac{\sqrt{30}}{21}\omega^{\mathbf{d}}_{22}+\frac{2\sqrt{10}}{21}\omega^{\mathbf{d}}_{42}-\frac{2\sqrt{70}}{21}\omega^{\mathbf{d}}_{44}$\\
\hline
$B_2$     &$22$  &$22$
          &$\omega^{\mathbf{d}}_{00}-\frac{2}{21}\omega^{\mathbf{d}}_{40}-\frac{\sqrt{5}}{7}\omega^{\mathbf{d}}_{20}-\frac{\sqrt{30}}{21}\omega^{\mathbf{d}}_{22}-\frac{2\sqrt{10}}{21}\omega^{\mathbf{d}}_{42}+\frac{2\sqrt{70}}{21}\omega^{\mathbf{d}}_{44}$\\
\hline
$E$       &$11$  &$11$   &$\omega^{\mathbf{d}}_{00}+\frac{\sqrt{5}}{10}\omega^{\mathbf{d}}_{20}$\\
          &$11$  &$22$   &$\frac{3\sqrt{5}i}{10}\omega^{\mathbf{d}}_{10}-\frac{\sqrt{14}i}{14}\omega^{\mathbf{d}}_{32}+\frac{\sqrt{105}i}{35}\omega^{\mathbf{d}}_{30}$\\
          &$22$  &$22$   &$\omega^{\mathbf{d}}_{00}+\frac{8}{21}\omega^{\mathbf{d}}_{44}+\frac{\sqrt{5}}{14}\omega^{\mathbf{d}}_{20}+\frac{2\sqrt{30}}{21}\omega^{\mathbf{d}}_{22}-\frac{\sqrt{10}}{7}\omega^{\mathbf{d}}_{42}$\\
\hline\hline
\end{tabular}
\end{center}
\end{table*}

\subsection{Phase shift formulas for baryon-baryon scattering in COM frame with twisted boundary condition}
\label{sec:baryon-baryon tbc}
In this subsection, we impose the twisted boundary condition for the two-particle system in the COM frame. The quantized momentum is
 $\bk=\frac{2\pi}{L}(\bn+\frac{{\boldsymbol{\theta}}}{2\pi})$ with $\bn$ the three-dimensional vector and $\boldsymbol{\theta}$ is the twist angle. By adjusting the twist angle, one can gain more momenta for two-particle energy levels.
L\"uscher's formula takes exactly the same form as shown in Eq.~(\ref{eq:luscher_general_form})
 except that all the labels of $\mathcal{M}^{\mathbf{c}(s)}_{JMl;J^{\prime}M^{\prime}l^{\prime}}$ are replaced
 by $\mathcal{M}^{\bphi(s)}_{JMl;J^{\prime}M^{\prime}l^{\prime}}$.
The explicit form of $\mathcal{M}^{\bphi(s)}_{JMl;J^{\prime}M^{\prime}l^{\prime}}$ is given by
 \beq
 \mathcal{M}^{\bphi(s)}_{JMl;J^{\prime}M^{\prime}l^{\prime}}(q^2,\eta)&=&\sum_{mm'\nu}\langle{J}M|lm;s\nu\rangle\langle{J^{\prime}}M^{\prime}|l^{\prime}m^{\prime};s\nu\rangle\nonumber\\
 & &\times{\mathcal{M}}^{\bphi}_{lm;l^{\prime}m^{\prime}}(q^2,\eta),
 \label{eq:M_JLM_tm}
 \eeq
 where
 \beq
 \mathcal{M}^{\bphi}_{lm;l^{\prime}m^{\prime}}(q^2,\eta)&=&\sum_{t=|l-l^{\prime}|}^{l+l^{\prime}}\sum_{t^{\prime}=-t}^{t}\frac{(-1)^{l}i^{l+l^{\prime}}}{\pi^{3/2}\eta{q^{t+1}}}\nonumber\\
 & &\times{Z}_{tt^{\prime}}^{\bphi}(q^2,\eta)\langle{l0t0}|{l^{\prime}0}\rangle\langle{lmtt^{\prime}}|{l^{\prime}m^{\prime}}\rangle\nonumber\\
 & &\times\sqrt{\frac{(2l+1)(2t+1)}{(2l^{\prime}+1)}},
 \label{M_m_tm}
 \eeq
 with
 \beq
 \label{eq:Z_jn_tm}
 Z^{\bphi}_{tt^{\prime}}(q^2,\eta)&=&\sum_{\mathbf{r}\in{\Gamma^{\bphi}}}\frac{\mathcal{Y}_{tt^{\prime}}(\mathbf{\tilde{n}})}
 {\mathbf{\tilde{n}}^2-q^2}\;.
 \eeq
 Here, $\Gamma^{\bphi}=\{\mathbf{\tilde{n}}\in\mathbbm{R}^3|\mathbf{\tilde{n}}=\mathbf{n}+(2\pi)^{-1}\bphi,\mathbf{n}\in\mathbbm{Z}^3\}$, and
 $\mathbbm{R}^3$ is the set of real 3-tuples. The the short-hand function for zeta function is
 \beq
\omega^{\bphi}_{lm}(q^2,\eta)=\frac{Z^{\bphi}_{lm}(q^2,\eta)}{\pi^{3/2}\eta{q^{l+1}}}.
\eeq
For the case with $\bphi=(0,0,\phi)$, the corresponding little group is written as $C_{4v}$. Taking the phase shift formula in the $A_1$ representation in Case A($s=s'=1$, $l=J\pm1$) as an example,
the phase shift formula is same with Eq.~(\ref{eq:A1caseamf}) except that all the labels of $\omega^{\mathbf{d}}_{lm}$ are replaced
 by $\omega^{\bphi}_{lm}$.
In particular, for $\bphi=(0,0,\pi)$, the corresponding little group is $D_{4h}$.
The phase shift formula is same with Eq.~(\ref{eq:A1-casea}) except that all the labels of $\omega_{lm}$ are replaced
 by $\omega^{\bphi}_{lm}$.

\section{Phase shift formulas for baryon-baryon scattering in COM frame in cubic box}
\label{sec:baryon-baryon cubic}
In this section, we briefly discuss phase shift formulas for baryon-baryon scattering in COM frame in cubic box.
On the one hand, we can perform the consistency checks and validation on the elongated results by comparing the two cases.
On the other hand, this can serve as a basis for exploring the relationship between the two cases, providing valuable insight into how the results transform from one to the other.

L\"uscher's formula takes the form given in Eq.~(\ref{eq:luscher_general_form}).
To write out a more explicit formula,
 we should consider the definite cubic symmetries. For the
 case of integer total momentum $J$, we need to consider the
 group of $O_h$, which contains $48$ elements and can be
 divided into $10$ conjugate classes: $A^{\pm}_{1}$, $A^{\pm}_{2}$, $E^{\pm}$, $T^{\pm}_{1}$ and $T^{\pm}_{2}$. For instance,
 for $J=0$, $1$,
 $2$, $\Lambda=2$, the decomposition into irreducible
 representation is given by $0^{\pm}=A_1^{\pm}$, $1^{\pm}=T_1^{\pm}$,
 $2^{\pm}=T_2^{\pm}\oplus{E^{\pm}}$ respectively~\cite{Lee:2017igf}.
 Then, we take L\"uscher's formula in Case A($s=s'=1$, $l=J\pm1$) as an example.
 Before writing out the explicit phase formula in this case, we list non-vanished matrix elements $\mathcal{M}^{\mathbf{c}(s)}_{Jl;J^{\prime}l^{\prime}}(\Gamma)$ in Tab.~\ref{tab:momentum_oh}.
\begin{table*}[!htb]
\begin{center}
\caption{$O_{h}$ symmetry group for angular momentum up to $J=2$ and $l=2$. The results are separated according to parity $(-1)^{J\pm1}$}
\label{tab:momentum_oh}       
\begin{tabular}{llll}
\hline\hline
$\Gamma$ &$Jl$  &$J^{\prime}l^{\prime}$ &$\mathcal{M}^{\mathbf{c}(1)}_{Jl;J^{\prime}l^{\prime}}(\Gamma)$\\
\noalign{\smallskip}\hline\noalign{\smallskip}
$A_1^{-}$ &$01$  &$01$
          &$\omega^{\mathbf{c}}_{00}-\frac{2\sqrt{30}}{15}\omega^{\mathbf{c}}_{22}$\\
          &$01$  &$21$
          &$-\frac{\sqrt{10}}{5}\omega^{\mathbf{c}}_{20}-\frac{2\sqrt{15}}{15}\omega^{\mathbf{c}}_{22}$\\
          &$21$  &$21$
          &$\omega^{\mathbf{c}}_{00}+\frac{\sqrt{5}}{5}\omega^{\mathbf{c}}_{20}-\frac{\sqrt{30}}{15}\omega^{\mathbf{c}}_{22}$\\
\hline
$E^{-}$   &$21$  &$21$
          &$\omega^{\mathbf{c}}_{00}-\frac{2\sqrt{30}}{15}\omega^{\mathbf{c}}_{22}$\\
\hline
$T_1^{+}$ &$10$  &$10$
          &$\omega^{\mathbf{c}}_{00}$\\
          &$10$  &$12$
          &$-\frac{2\sqrt{15}}{15}\omega^{\mathbf{c}}_{22}$\\
          &$12$  &$12$
          &$\omega^{\mathbf{c}}_{00}-\frac{\sqrt{30}}{15}\omega^{\mathbf{c}}_{22}$\\
\hline
$T_2^{-}$ &$21$  &$21$
          &$\omega^{\mathbf{c}}_{00}+\frac{\sqrt{30}}{15}\omega^{\mathbf{c}}_{22}$\\
\hline\hline
\end{tabular}
\end{center}
\end{table*}
According to the non-zero matrix elements, one can obtain phase shift formula in the definite symmetry. If we focus in $A_1^-$ representation, the phase shift formula is Eq.~(\ref{eq:A1-cubic}).
\bw
\beq
\left|
\ba
 {cc}
\cot\delta_{01}-(\omega^{\mathbf{c}}_{00}-\frac{2\sqrt{30}}{15}\omega^{\mathbf{c}}_{22})   &-\frac{\sqrt{10}}{5}\omega^{\mathbf{c}}_{20}-\frac{2\sqrt{15}}{15}\omega^{\mathbf{c}}_{22}\\
-\frac{\sqrt{10}}{5}\omega^{\mathbf{c}}_{20}-\frac{2\sqrt{15}}{15}\omega^{\mathbf{c}}_{22}     &\cot\delta_{21}-(\omega^{\mathbf{c}}_{00}+\frac{\sqrt{5}}{5}\omega^{\mathbf{c}}_{20}-\frac{\sqrt{30}}{15}\omega^{\mathbf{c}}_{22})
\ea
\right|=0
 \label{eq:A1-cubic}
\eeq
\ew
If we ignore the mixing with $J=2$, one can extract S-wave phase shift from the following form:
\be
\cot\delta_{01}=\omega^{\mathbf{c}}_{00}-\frac{2\sqrt{30}}{15}\omega^{\mathbf{c}}_{22}.
\ee
If we consider D-wave resonance, we can give the phase shift formulas in $E^-$ and $T_2^-$ representations respectively, i.e.,
\be
\cot\delta_{21}=\omega^{\mathbf{c}}_{00}-\frac{2\sqrt{30}}{15}\omega^{\mathbf{c}}_{22},
\ee
and
\be
\cot\delta_{21}=\omega^{\mathbf{c}}_{00}+\frac{\sqrt{30}}{15}\omega^{\mathbf{c}}_{22}.
\ee
Finally, for the $T_1^+$ representation with $O_{h}$ group, the
phase shift formula is given by Eq.~(\ref{eq:T1+cubic}).
\bw
\be
\left|
\left(
S_{2\times2}-I_{2\times2}
\right)
\left(
\ba
{cc}
\omega^{\mathbf{c}}_{00}   &-\frac{2\sqrt{15}}{15}\omega^{\mathbf{c}}_{22}\\
-\frac{2\sqrt{15}}{15}\omega^{\mathbf{c}}_{22}       &\omega^{\mathbf{c}}_{00}-\frac{\sqrt{30}}{15}\omega^{\mathbf{c}}_{22}
\ea
\right)
 -i\left(
 S_{2\times2}+I_{2\times2}
\right)
\right|
 =0
 \label{eq:T1+cubic}
\ee
\ew

Then, let us discuss the relationship of two-particle scattering L\"uscher's formula between the cubic case and the elongated case.
In Tab.~{\ref{tab:relationship}}, we listed the symmetry relationship between $O_h$ and $D_{4h}$ \cite{Lee:2017igf}.
\begin{table*}[!htb]
\begin{center}
\caption{Subduction rules in the descent in symmetry in the group from cubic box($O_h$) to the elongated box($D_{4h}$). }
\label{tab:relationship}       
\begin{tabular}{lllllllllll}
\hline\hline
$O_h$ &$A_1^+$ &$A_2^+$ &$E^+$ &$T_1^+$  &$T_2^+$ &$A_1^-$ &$A_2^-$ &$E^-$ &$T_1^-$  &$T_2^-$\\
\noalign{\smallskip}\hline\noalign{\smallskip}
$D_{4h}$ &$A_1^+$ &$B_1^+$ &$A_1^+\oplus{B_1^+}$ &$A_2^+\oplus{E^+}$  &$B_2^+\oplus{E^+}$ &$A_1^-$ &$B_1^-$ &$A_1^-\oplus{B_1^-}$ &$A_2^-\oplus{E^-}$  &$B_2^-\oplus{E^-}$\\
\hline\hline
\end{tabular}
\end{center}
\end{table*}
By using the relationship in Tab.~{\ref{tab:relationship}}, one can readily obtain the following relationships.
For $(J=J^{\prime}=0)$, the $A_1^-$ has one-to-one correspondence:
\be
\mathcal{M}^{\mathbf{c}(1)}_{01;01}(A_1^-)=\mathcal{M}^{(1)}_{01;01}(A_1^-).
\label{eq:A1corr}
\ee
Then, for $(J=J^{\prime}=2)$, the $E^-$ splits into $A_1^-$ and $B_1^-$, i.e.,
\be
\mathcal{M}^{\mathbf{c}(1)}_{21;21}(E^-)=\frac{1}{2}\mathcal{M}^{(1)}_{21;21}(A_1^-)+\frac{1}{2}\mathcal{M}^{(1)}_{21;21}(B_1^-).
\label{eq:Ecorr}
\ee
Next, for $(J=J^{\prime}=1)$, the $T_1^+$ is divided into $A_2^+$ and $E^+$ so that
\be
 \left\{ \begin{aligned}
 \mathcal{M}^{\mathbf{c}(1)}_{10;10}(T_1^+)&=\frac{1}{3}\mathcal{M}^{(1)}_{10;10}(A_2^+)+\frac{2}{3}\mathcal{M}^{(1)}_{10;10}(E^+)\\
 \mathcal{M}^{\mathbf{c}(1)}_{10;12}(T_1^+)&=\frac{1}{3}\mathcal{M}^{(1)}_{10;12}(A_2^+)+\frac{2}{3}\mathcal{M}^{(1)}_{10;12}(E^+)\\
 \mathcal{M}^{\mathbf{c}(1)}_{12;12}(T_1^+)&=\frac{1}{3}\mathcal{M}^{(1)}_{12;12}(A_2^+)+\frac{2}{3}\mathcal{M}^{(1)}_{12;12}(E^+)\\
 \end{aligned} \right.\;.
 \ee
Then, for $(J=J^{\prime}=2)$, the $T_2^-$ splits into $B_2^-$ and $E^-$, and we have
\be
\mathcal{M}^{\mathbf{c}(1)}_{21;21}(T_2^-)=\frac{1}{3}\mathcal{M}^{(1)}_{21;21}(B_2^-)+\frac{2}{3}\mathcal{M}^{(1)}_{21;21}(E^-)
\label{eq:T2corr}
\ee
Finally, we discuss the relationship of phase shift formulas between the cubic boxes and the elongated boxes. Here, we take $T_2^-$ representation in cubic boxes as an example.
From the elongated box to the cubic symmetry, the matrix elements $\mathcal{M}^{(1)}_{21;21}(B_2^-)$ and $\mathcal{M}^{(1)}_{21;21}(E^-)$  will individually approach $\mathcal{M}^{\mathbf{c}(1)}_{21;21}(T_2^-)$ in the limit $\eta=1$. From Eq.~(\ref{eq:T2corr}), one can see
 how to follow this limit by subduction rule in this particular channel. The relationship translates directly into one for the phase shift as follows:
\beq
\cot\delta_{21}(T_2^-)=\frac{1}{3}\cot\delta_{21}(B_2^-)+\frac{2}{3}\cot\delta_{21}(E^-).
\eeq

\section{Discussions and conclusions}
 \label{sec:conclude}
In this paper, we have derived L\"uscher phase shift formulas for baryon-baryon elastic scattering in elongated boxes. We show the cases where the baryon-baryon state is in COM frame with respect to the box, or moving, or with twisted boundary condition along the elongated direction.
As a consistency check and validation on the elongated results, we have also derived the results in the cubic box by using the same approach.
There are two differences between these two cases. One is the symmetry of the two-particle system and the other is the matrix $\mathcal{M}$.
We take the two-baryon scattering in COM frame as an example. In the cubic case, the symmetry group is $O_h$ and in the elongated case, the symmetry group becomes $D_{4h}$.
In addition, the factor of $\eta$ in matrix $\mathcal{M}$ in the elongated box is equal to $1$ for the cubic case. Our interest in elongated boxes stems from the fact that they allow us to vary the geometry of the box, and consequently the kinematics, with minimal amount of computer resources. On the other hand, elongated boxes have a different symmetry group than the cubic case and this has to be taken into account when designing interpolators and connecting the infinite volume phase-shifts with the two-body energies.

Let us discuss some possible applications of phase shift formula derived in this paper.
Some typical examples for scattering with two spin-$1/2$  particles are listed here, and all examples are highly relevant
in the study of the composition of dense nuclear matter
which forms the neutron stars.
One typical
example is the $\Lambda-\Lambda$, $N-\Xi$, $\Sigma-\Sigma$ scattering, where the existence of H dibaryon state has been discussed as the remaining of bound state in
flavor singlet channel.
Another example is $\Lambda-N$ and $\Sigma-N$ scattering which is useful to study properties of hyperonic matters inside the
neutron stars.
$\Xi-\Xi$, $N-N$ and $N\Sigma$ scattering and $N\Lambda$ scattering follow.
These examples have been studied by using HAL QCD method in Refs.~\cite{Sasaki:2017ysy,Nemura:2017bbw,Doi:2017cfx,Ishii:2017xud}. In principle, these issues can also be
studied with L\"uscher's formulas in detail by using lattice QCD simulations in elongated boxes.

To summarize, in this paper we have generalized two-particle scattering phase shift
formulas to the case of particles with spin $1/2$, below the
inelastic threshold in the COM frame and MF in the elongated box and COM frame in cubic box, respectively.
Using a quantum mechanical model, a relation between the energy of the two-particle
system and phase shift is established. It is verified
that the phase shift formulas in elongated box in the limit $\eta=1$ and cubic box are consistent.  Although we focus on  the
scattering between two particles
with all of the spin being $1/2$, there are not essential difficulties
in generalizing it to cases with arbitrary spin in any
number of channels. We expect that these relations will be helpful for the
study of baryon-baryon scattering in lattice QCD simulations.

\section{Acknowledgments}
The authors would like to thank Xu Feng for helpful discussions.
This work is supported by the National Science Foundation of China (NSFC) under the grant No.
11505132, 11504285, 11705072 and the Scientific Research Program Funded by Natural Science Basic Research Plan in
Shaanxi Province of China (Program No.~2016JQ1009) and supported By Young Talent fund of University Association
for Science and Technology in Shaanxi, China (Program No.~20170608). This work is also supported by the Fundamental
Research Funds for the Central Universities of Lanzhou University under Grants 223000-862637.

 \bibliographystyle{apsrev4-1}

\begin{thebibliography}{10}%
\makeatletter
\providecommand \@ifxundefined [1]{%
 \ifx #1\undefined \expandafter \@firstoftwo
 \else \expandafter \@secondoftwo
\fi
}%
\providecommand \@ifnum [1]{%
 \ifnum #1\expandafter \@firstoftwo
 \else \expandafter \@secondoftwo
\fi
}%
\providecommand \enquote [1]{``#1''}%
\providecommand \bibnamefont  [1]{#1}%
\providecommand \bibfnamefont [1]{#1}%
\providecommand \citenamefont [1]{#1}%
\providecommand\href[0]{\@sanitize\@href}%
\providecommand\@href[1]{\endgroup\@@startlink{#1}\endgroup\@@href}%
\providecommand\@@href[1]{#1\@@endlink}%
\providecommand \@sanitize [0]{\begingroup\catcode`\&12\catcode`\#12\relax}%
\@ifxundefined \pdfoutput {\@firstoftwo}{%
 \@ifnum{\z@=\pdfoutput}{\@firstoftwo}{\@secondoftwo}%
}{%
 \providecommand\@@startlink[1]{\leavevmode\special{html:<a href="#1">}}%
 \providecommand\@@endlink[0]{\special{html:</a>}}%
}{%
 \providecommand\@@startlink[1]{%
  \leavevmode
  \pdfstartlink
   attr{/Border[0 0 1 ]/H/I/C[0 1 1]}%
   user{/Subtype/Link/A<</Type/Action/S/URI/URI(#1)>>}%
  \relax
 }%
 \providecommand\@@endlink[0]{\pdfendlink}%
}%
\providecommand \url  [0]{\begingroup\@sanitize \@url }%
\providecommand \@url [1]{\endgroup\@href {#1}{\urlprefix}}%
\providecommand \urlprefix [0]{URL }%
\providecommand \Eprint[0]{\href }%
\@ifxundefined \urlstyle {%
  \providecommand \doi [1]{doi:\discretionary{}{}{}#1}%
}{%
  \providecommand \doi [0]{doi:\discretionary{}{}{}\begingroup
  \urlstyle{rm}\Url }%
}%
\providecommand \doibase [0]{http://dx.doi.org/}%
\providecommand \Doi[1]{\href{\doibase#1}}%
\providecommand \bibAnnote [3]{%
  \BibitemShut{#1}%
  \begin{quotation}\noindent
    \textsc{Key:}\ #2\\\textsc{Annotation:}\ #3%
  \end{quotation}%
}%
\providecommand \bibAnnoteFile [2]{%
  \IfFileExists{#2}{\bibAnnote {#1} {#2} {\input{#2}}}{}%
}%
\providecommand \typeout [0]{\immediate \write \m@ne }%
\providecommand \selectlanguage [0]{\@gobble}%
\providecommand \bibinfo [0]{\@secondoftwo}%
\providecommand \bibfield [0]{\@secondoftwo}%
\providecommand \translation [1]{[#1]}%
\providecommand \BibitemOpen[0]{}%
\providecommand \bibitemStop [0]{}%
\providecommand \bibitemNoStop [0]{.\EOS\space}%
\providecommand \EOS [0]{\spacefactor3000\relax}%
\providecommand \BibitemShut [1]{\csname bibitem#1\endcsname}%
\bibitem{Wilson:2014cna}%
  \BibitemOpen
  \bibfield{author}{%
  \bibinfo {author} {\bibfnamefont{D.~J.}\ \bibnamefont{Wilson}}, \bibinfo
  {author} {\bibfnamefont{J.~J.}\ \bibnamefont{Dudek}}, \bibinfo {author}
  {\bibfnamefont{R.~G.}\ \bibnamefont{Edwards}},\ and\ \bibinfo {author}
  {\bibfnamefont{C.~E.}\ \bibnamefont{Thomas}},\ }%
  \bibfield{journal}{%
  \Doi{10.1103/PhysRevD.91.054008}{\bibinfo {journal} {Phys. Rev.}}\ }%
  \textbf{\bibinfo {volume} {D91}},\ \bibinfo {pages} {054008} (\bibinfo {year}
  {2015}),\ \Eprint{http://arxiv.org/abs/1411.2004}{arXiv:1411.2004 [hep-ph]}%
  \bibAnnoteFile{NoStop}{Wilson:2014cna}%
\bibitem{Wilson:2015dqa}%
  \BibitemOpen
  \bibfield{author}{%
  \bibinfo {author} {\bibfnamefont{D.~J.}\ \bibnamefont{Wilson}}, \bibinfo
  {author} {\bibfnamefont{R.~A.}\ \bibnamefont{Briceno}}, \bibinfo {author}
  {\bibfnamefont{J.~J.}\ \bibnamefont{Dudek}}, \bibinfo {author}
  {\bibfnamefont{R.~G.}\ \bibnamefont{Edwards}},\ and\ \bibinfo {author}
  {\bibfnamefont{C.~E.}\ \bibnamefont{Thomas}},\ }%
  \bibfield{journal}{%
  \Doi{10.1103/PhysRevD.92.094502}{\bibinfo {journal} {Phys. Rev.}}\ }%
  \textbf{\bibinfo {volume} {D92}},\ \bibinfo {pages} {094502} (\bibinfo {year}
  {2015}),\ \Eprint{http://arxiv.org/abs/1507.02599}{arXiv:1507.02599
  [hep-ph]}%
  \bibAnnoteFile{NoStop}{Wilson:2015dqa}%
\bibitem{Dudek:2016cru}%
  \BibitemOpen
  \bibfield{author}{%
  \bibinfo {author} {\bibfnamefont{J.~J.}\ \bibnamefont{Dudek}}, \bibinfo
  {author} {\bibfnamefont{R.~G.}\ \bibnamefont{Edwards}},\ and\ \bibinfo
  {author} {\bibfnamefont{D.~J.}\ \bibnamefont{Wilson}} (\bibinfo
  {collaboration} {Hadron Spectrum}),\ }%
  \bibfield{journal}{%
  \Doi{10.1103/PhysRevD.93.094506}{\bibinfo {journal} {Phys. Rev.}}\ }%
  \textbf{\bibinfo {volume} {D93}},\ \bibinfo {pages} {094506} (\bibinfo {year}
  {2016}),\ \Eprint{http://arxiv.org/abs/1602.05122}{arXiv:1602.05122
  [hep-ph]}%
  \bibAnnoteFile{NoStop}{Dudek:2016cru}%
\bibitem{Aoki:2007rd}%
  \BibitemOpen
  \bibfield{author}{%
  \bibinfo {author} {\bibfnamefont{S.}~\bibnamefont{Aoki}} \emph{et~al.}
  (\bibinfo {collaboration} {CP-PACS}),\ }%
  \bibfield{journal}{%
  \Doi{10.1103/PhysRevD.76.094506}{\bibinfo {journal} {Phys. Rev.}}\ }%
  \textbf{\bibinfo {volume} {D76}},\ \bibinfo {pages} {094506} (\bibinfo {year}
  {2007}),\ \Eprint{http://arxiv.org/abs/0708.3705}{arXiv:0708.3705 [hep-lat]}%
  \bibAnnoteFile{NoStop}{Aoki:2007rd}%
\bibitem{Feng:2010es}%
  \BibitemOpen
  \bibfield{author}{%
  \bibinfo {author} {\bibfnamefont{X.}~\bibnamefont{Feng}}, \bibinfo {author}
  {\bibfnamefont{K.}~\bibnamefont{Jansen}},\ and\ \bibinfo {author}
  {\bibfnamefont{D.~B.}\ \bibnamefont{Renner}},\ }%
  \bibfield{journal}{%
  \Doi{10.1103/PhysRevD.83.094505}{\bibinfo {journal} {Phys. Rev.}}\ }%
  \textbf{\bibinfo {volume} {D83}},\ \bibinfo {pages} {094505} (\bibinfo {year}
  {2011}),\ \Eprint{http://arxiv.org/abs/1011.5288}{arXiv:1011.5288 [hep-lat]}%
  \bibAnnoteFile{NoStop}{Feng:2010es}%
\bibitem{Lang:2011mn}%
  \BibitemOpen
  \bibfield{author}{%
  \bibinfo {author} {\bibfnamefont{C.~B.}\ \bibnamefont{Lang}}, \bibinfo
  {author} {\bibfnamefont{D.}~\bibnamefont{Mohler}}, \bibinfo {author}
  {\bibfnamefont{S.}~\bibnamefont{Prelovsek}},\ and\ \bibinfo {author}
  {\bibfnamefont{M.}~\bibnamefont{Vidmar}},\ }%
  \bibfield{journal}{%
  \Doi{10.1103/PhysRevD.89.059903, 10.1103/PhysRevD.84.054503}{\bibinfo
  {journal} {Phys. Rev.}}\ }%
  \textbf{\bibinfo {volume} {D84}},\ \bibinfo {pages} {054503} (\bibinfo {year}
  {2011}),\ \bibinfo {note} {[Erratum: Phys. Rev.D89,no.5,059903(2014)]},\
  \Eprint{http://arxiv.org/abs/1105.5636}{arXiv:1105.5636 [hep-lat]}%
  \bibAnnoteFile{NoStop}{Lang:2011mn}%
\bibitem{Feng:2014gba}%
  \BibitemOpen
  \bibfield{author}{%
  \bibinfo {author} {\bibfnamefont{X.}~\bibnamefont{Feng}}, \bibinfo {author}
  {\bibfnamefont{S.}~\bibnamefont{Aoki}}, \bibinfo {author}
  {\bibfnamefont{S.}~\bibnamefont{Hashimoto}},\ and\ \bibinfo {author}
  {\bibfnamefont{T.}~\bibnamefont{Kaneko}},\ }%
  \bibfield{journal}{%
  \Doi{10.1103/PhysRevD.91.054504}{\bibinfo {journal} {Phys. Rev.}}\ }%
  \textbf{\bibinfo {volume} {D91}},\ \bibinfo {pages} {054504} (\bibinfo {year}
  {2015}),\ \Eprint{http://arxiv.org/abs/1412.6319}{arXiv:1412.6319 [hep-lat]}%
  \bibAnnoteFile{NoStop}{Feng:2014gba}%
\bibitem{Briceno:2016mjc}%
  \BibitemOpen
  \bibfield{author}{%
  \bibinfo {author} {\bibfnamefont{R.~A.}\ \bibnamefont{Briceno}}, \bibinfo
  {author} {\bibfnamefont{J.~J.}\ \bibnamefont{Dudek}}, \bibinfo {author}
  {\bibfnamefont{R.~G.}\ \bibnamefont{Edwards}},\ and\ \bibinfo {author}
  {\bibfnamefont{D.~J.}\ \bibnamefont{Wilson}},\ }%
  \bibfield{journal}{%
  \Doi{10.1103/PhysRevLett.118.022002}{\bibinfo {journal} {Phys. Rev. Lett.}}\
  }%
  \textbf{\bibinfo {volume} {118}},\ \bibinfo {pages} {022002} (\bibinfo {year}
  {2017}),\ \Eprint{http://arxiv.org/abs/1607.05900}{arXiv:1607.05900
  [hep-ph]}%
  \bibAnnoteFile{NoStop}{Briceno:2016mjc}%
\bibitem{Sun:2014aya}%
  \BibitemOpen
  \bibfield{author}{%
  \bibinfo {author} {\bibfnamefont{Z.-F.}\ \bibnamefont{Sun}}, \bibinfo
  {author} {\bibfnamefont{Z.-W.}\ \bibnamefont{Liu}}, \bibinfo {author}
  {\bibfnamefont{X.}~\bibnamefont{Liu}},\ and\ \bibinfo {author}
  {\bibfnamefont{S.-L.}\ \bibnamefont{Zhu}},\ }%
  \bibfield{journal}{%
  \Doi{10.1103/PhysRevD.91.094030}{\bibinfo {journal} {Phys. Rev.}}\ }%
  \textbf{\bibinfo {volume} {D91}},\ \bibinfo {pages} {094030} (\bibinfo {year}
  {2015}),\ \Eprint{http://arxiv.org/abs/1411.2117}{arXiv:1411.2117 [hep-ph]}%
  \bibAnnoteFile{NoStop}{Sun:2014aya}%
\bibitem{Liu:2015ktc}%
  \BibitemOpen
  \bibfield{author}{%
  \bibinfo {author} {\bibfnamefont{Z.-W.}\ \bibnamefont{Liu}}, \bibinfo
  {author} {\bibfnamefont{W.}~\bibnamefont{Kamleh}}, \bibinfo {author}
  {\bibfnamefont{D.~B.}\ \bibnamefont{Leinweber}}, \bibinfo {author}
  {\bibfnamefont{F.~M.}\ \bibnamefont{Stokes}}, \bibinfo {author}
  {\bibfnamefont{A.~W.}\ \bibnamefont{Thomas}},\ and\ \bibinfo {author}
  {\bibfnamefont{J.-J.}\ \bibnamefont{Wu}},\ }%
  \bibfield{journal}{%
  \Doi{10.1103/PhysRevLett.116.082004}{\bibinfo {journal} {Phys. Rev. Lett.}}\
  }%
  \textbf{\bibinfo {volume} {116}},\ \bibinfo {pages} {082004} (\bibinfo {year}
  {2016}),\ \Eprint{http://arxiv.org/abs/1512.00140}{arXiv:1512.00140
  [hep-lat]}%
  \bibAnnoteFile{NoStop}{Liu:2015ktc}%
\bibitem{Liu:2016uzk}%
  \BibitemOpen
  \bibfield{author}{%
  \bibinfo {author} {\bibfnamefont{Z.-W.}\ \bibnamefont{Liu}}, \bibinfo
  {author} {\bibfnamefont{W.}~\bibnamefont{Kamleh}}, \bibinfo {author}
  {\bibfnamefont{D.~B.}\ \bibnamefont{Leinweber}}, \bibinfo {author}
  {\bibfnamefont{F.~M.}\ \bibnamefont{Stokes}}, \bibinfo {author}
  {\bibfnamefont{A.~W.}\ \bibnamefont{Thomas}},\ and\ \bibinfo {author}
  {\bibfnamefont{J.-J.}\ \bibnamefont{Wu}},\ }%
  \bibfield{journal}{%
  \Doi{10.1103/PhysRevD.95.034034}{\bibinfo {journal} {Phys. Rev.}}\ }%
  \textbf{\bibinfo {volume} {D95}},\ \bibinfo {pages} {034034} (\bibinfo {year}
  {2017}),\ \Eprint{http://arxiv.org/abs/1607.04536}{arXiv:1607.04536
  [nucl-th]}%
  \bibAnnoteFile{NoStop}{Liu:2016uzk}%
\bibitem{Liu:2016wxq}%
  \BibitemOpen
  \bibfield{author}{%
  \bibinfo {author} {\bibfnamefont{Z.-W.}\ \bibnamefont{Liu}}, \bibinfo
  {author} {\bibfnamefont{J.~M.~M.}\ \bibnamefont{Hall}}, \bibinfo {author}
  {\bibfnamefont{D.~B.}\ \bibnamefont{Leinweber}}, \bibinfo {author}
  {\bibfnamefont{A.~W.}\ \bibnamefont{Thomas}},\ and\ \bibinfo {author}
  {\bibfnamefont{J.-J.}\ \bibnamefont{Wu}},\ }%
  \bibfield{journal}{%
  \Doi{10.1103/PhysRevD.95.014506}{\bibinfo {journal} {Phys. Rev.}}\ }%
  \textbf{\bibinfo {volume} {D95}},\ \bibinfo {pages} {014506} (\bibinfo {year}
  {2017}),\ \Eprint{http://arxiv.org/abs/1607.05856}{arXiv:1607.05856
  [nucl-th]}%
  \bibAnnoteFile{NoStop}{Liu:2016wxq}%
\bibitem{Kiratidis:2016hda}%
  \BibitemOpen
  \bibfield{author}{%
  \bibinfo {author} {\bibfnamefont{A.~L.}\ \bibnamefont{Kiratidis}}, \bibinfo
  {author} {\bibfnamefont{W.}~\bibnamefont{Kamleh}}, \bibinfo {author}
  {\bibfnamefont{D.~B.}\ \bibnamefont{Leinweber}}, \bibinfo {author}
  {\bibfnamefont{Z.-W.}\ \bibnamefont{Liu}}, \bibinfo {author}
  {\bibfnamefont{F.~M.}\ \bibnamefont{Stokes}},\ and\ \bibinfo {author}
  {\bibfnamefont{A.~W.}\ \bibnamefont{Thomas}},\ }%
  \bibfield{journal}{%
  \Doi{10.1103/PhysRevD.95.074507}{\bibinfo {journal} {Phys. Rev.}}\ }%
  \textbf{\bibinfo {volume} {D95}},\ \bibinfo {pages} {074507} (\bibinfo {year}
  {2017}),\ \Eprint{http://arxiv.org/abs/1608.03051}{arXiv:1608.03051
  [hep-lat]}%
  \bibAnnoteFile{NoStop}{Kiratidis:2016hda}%
\bibitem{luscher86:finiteb}%
  \BibitemOpen
  \bibfield{author}{%
  \bibinfo {author} {\bibfnamefont{M.}~\bibnamefont{L{\"u}scher}},\ }%
  \bibfield{journal}{%
  \Doi{10.1007/BF01211097}{\bibinfo {journal} {Commun. Math. Phys.}}\ }%
  \textbf{\bibinfo {volume} {105}},\ \bibinfo {pages} {153} (\bibinfo {year}
  {1986})%
  \bibAnnoteFile{NoStop}{luscher86:finiteb}%
\bibitem{luscher90:finite}%
  \BibitemOpen
  \bibfield{author}{%
  \bibinfo {author} {\bibfnamefont{M.}~\bibnamefont{L{\"u}scher}}\ and\
  \bibinfo {author} {\bibfnamefont{U.}~\bibnamefont{Wolff}},\ }%
  \bibfield{journal}{%
  \Doi{10.1016/0550-3213(90)90540-T}{\bibinfo {journal} {Nucl. Phys.}}\ }%
  \textbf{\bibinfo {volume} {B339}},\ \bibinfo {pages} {222} (\bibinfo {year}
  {1990})%
  \bibAnnoteFile{NoStop}{luscher90:finite}%
\bibitem{luscher91:finitea}%
  \BibitemOpen
  \bibfield{author}{%
  \bibinfo {author} {\bibfnamefont{M.}~\bibnamefont{L{\"u}scher}},\ }%
  \bibfield{journal}{%
  \Doi{10.1016/0550-3213(91)90366-6}{\bibinfo {journal} {Nucl. Phys.}}\ }%
  \textbf{\bibinfo {volume} {B354}},\ \bibinfo {pages} {531} (\bibinfo {year}
  {1991})%
  \bibAnnoteFile{NoStop}{luscher91:finitea}%
\bibitem{luscher91:finiteb}%
  \BibitemOpen
  \bibfield{author}{%
  \bibinfo {author} {\bibfnamefont{M.}~\bibnamefont{L{\"u}scher}},\ }%
  \bibfield{journal}{%
  \Doi{10.1016/0550-3213(91)90584-K}{\bibinfo {journal} {Nucl. Phys.}}\ }%
  \textbf{\bibinfo {volume} {B364}},\ \bibinfo {pages} {237} (\bibinfo {year}
  {1991})%
  \bibAnnoteFile{NoStop}{luscher91:finiteb}%
\bibitem{Gupta:1993rn}%
  \BibitemOpen
  \bibfield{author}{%
  \bibinfo {author} {\bibfnamefont{R.}~\bibnamefont{Gupta}}, \bibinfo {author}
  {\bibfnamefont{A.}~\bibnamefont{Patel}},\ and\ \bibinfo {author}
  {\bibfnamefont{S.~R.}\ \bibnamefont{Sharpe}},\ }%
  \bibfield{journal}{%
  \Doi{10.1103/PhysRevD.48.388}{\bibinfo {journal} {Phys. Rev. D}}\ }%
  \textbf{\bibinfo {volume} {48}},\ \bibinfo {pages} {388} (\bibinfo {year}
  {1993}),\ \Eprint{http://arxiv.org/abs/hep-lat/9301016}{arXiv:hep-lat/9301016
  [hep-lat]}%
  \bibAnnoteFile{NoStop}{Gupta:1993rn}%
\bibitem{Fukugita:1994ve}%
  \BibitemOpen
  \bibfield{author}{%
  \bibinfo {author} {\bibfnamefont{M.}~\bibnamefont{Fukugita}}, \bibinfo
  {author} {\bibfnamefont{Y.}~\bibnamefont{Kuramashi}}, \bibinfo {author}
  {\bibfnamefont{M.}~\bibnamefont{Okawa}}, \bibinfo {author}
  {\bibfnamefont{H.}~\bibnamefont{Mino}},\ and\ \bibinfo {author}
  {\bibfnamefont{A.}~\bibnamefont{Ukawa}},\ }%
  \bibfield{journal}{%
  \Doi{10.1103/PhysRevD.52.3003}{\bibinfo {journal} {Phys. Rev. D}}\ }%
  \textbf{\bibinfo {volume} {52}},\ \bibinfo {pages} {3003} (\bibinfo {year}
  {1995}),\ \Eprint{http://arxiv.org/abs/hep-lat/9501024}{arXiv:hep-lat/9501024
  [hep-lat]}%
  \bibAnnoteFile{NoStop}{Fukugita:1994ve}%
\bibitem{Aoki:1999pt}%
  \BibitemOpen
  \bibfield{author}{%
  \bibinfo {author} {\bibfnamefont{S.}~\bibnamefont{Aoki}} \emph{et~al.}
  (\bibinfo {collaboration} {JLQCD Collaboration}),\ }%
  \bibfield{journal}{%
  \bibinfo {journal} {Nucl. Phys. Proc. Suppl.}\ }%
  \textbf{\bibinfo {volume} {83}},\ \bibinfo {pages} {241} (\bibinfo {year}
  {2000}),\ \Eprint{http://arxiv.org/abs/hep-lat/9911025}{arXiv:hep-lat/9911025
  [hep-lat]}%
  \bibAnnoteFile{NoStop}{Aoki:1999pt}%
\bibitem{Aoki:2002in}%
  \BibitemOpen
  \bibfield{author}{%
  \bibinfo {author} {\bibfnamefont{S.}~\bibnamefont{Aoki}} \emph{et~al.}
  (\bibinfo {collaboration} {JLQCD Collaboration}),\ }%
  \bibfield{journal}{%
  \Doi{10.1103/PhysRevD.66.077501}{\bibinfo {journal} {Phys. Rev. D}}\ }%
  \textbf{\bibinfo {volume} {66}},\ \bibinfo {pages} {077501} (\bibinfo {year}
  {2002}),\ \Eprint{http://arxiv.org/abs/hep-lat/0206011}{arXiv:hep-lat/0206011
  [hep-lat]}%
  \bibAnnoteFile{NoStop}{Aoki:2002in}%
\bibitem{Liu:2001ss}%
  \BibitemOpen
  \bibfield{author}{%
  \bibinfo {author} {\bibfnamefont{C.}~\bibnamefont{Liu}}, \bibinfo {author}
  {\bibfnamefont{J.-h.}\ \bibnamefont{Zhang}}, \bibinfo {author}
  {\bibfnamefont{Y.}~\bibnamefont{Chen}},\ and\ \bibinfo {author}
  {\bibfnamefont{J.}~\bibnamefont{Ma}},\ }%
  \bibfield{journal}{%
  \Doi{10.1016/S0550-3213(01)00662-9}{\bibinfo {journal} {Nucl. Phys. B}}\ }%
  \textbf{\bibinfo {volume} {624}},\ \bibinfo {pages} {360} (\bibinfo {year}
  {2002}),\ \Eprint{http://arxiv.org/abs/hep-lat/0109020}{arXiv:hep-lat/0109020
  [hep-lat]}%
  \bibAnnoteFile{NoStop}{Liu:2001ss}%
\bibitem{Du:2004ib}%
  \BibitemOpen
  \bibfield{author}{%
  \bibinfo {author} {\bibfnamefont{X.}~\bibnamefont{Du}}, \bibinfo {author}
  {\bibfnamefont{G.-w.}\ \bibnamefont{Meng}}, \bibinfo {author}
  {\bibfnamefont{C.}~\bibnamefont{Miao}},\ and\ \bibinfo {author}
  {\bibfnamefont{C.}~\bibnamefont{Liu}},\ }%
  \bibfield{journal}{%
  \Doi{10.1142/S0217751X04019573}{\bibinfo {journal} {Int. J. Mod. Phys. A}}\
  }%
  \textbf{\bibinfo {volume} {19}},\ \bibinfo {pages} {5609} (\bibinfo {year}
  {2004}),\ \Eprint{http://arxiv.org/abs/hep-lat/0404017}{arXiv:hep-lat/0404017
  [hep-lat]}%
  \bibAnnoteFile{NoStop}{Du:2004ib}%
\bibitem{Aoki:2005uf}%
  \BibitemOpen
  \bibfield{author}{%
  \bibinfo {author} {\bibfnamefont{S.}~\bibnamefont{Aoki}} \emph{et~al.}
  (\bibinfo {collaboration} {CP-PACS Collaboration}),\ }%
  \bibfield{journal}{%
  \Doi{10.1103/PhysRevD.71.094504}{\bibinfo {journal} {Phys. Rev. D}}\ }%
  \textbf{\bibinfo {volume} {71}},\ \bibinfo {pages} {094504} (\bibinfo {year}
  {2005}),\ \Eprint{http://arxiv.org/abs/hep-lat/0503025}{arXiv:hep-lat/0503025
  [hep-lat]}%
  \bibAnnoteFile{NoStop}{Aoki:2005uf}%
\bibitem{Aoki:2002ny}%
  \BibitemOpen
  \bibfield{author}{%
  \bibinfo {author} {\bibfnamefont{S.}~\bibnamefont{Aoki}} \emph{et~al.}
  (\bibinfo {collaboration} {CP-PACS Collaboration}),\ }%
  \bibfield{journal}{%
  \Doi{10.1103/PhysRevD.67.014502}{\bibinfo {journal} {Phys. Rev. D}}\ }%
  \textbf{\bibinfo {volume} {67}},\ \bibinfo {pages} {014502} (\bibinfo {year}
  {2003}),\ \Eprint{http://arxiv.org/abs/hep-lat/0209124}{arXiv:hep-lat/0209124
  [hep-lat]}%
  \bibAnnoteFile{NoStop}{Aoki:2002ny}%
\bibitem{Yamazaki:2004qb}%
  \BibitemOpen
  \bibfield{author}{%
  \bibinfo {author} {\bibfnamefont{T.}~\bibnamefont{Yamazaki}} \emph{et~al.}
  (\bibinfo {collaboration} {CP-PACS Collaboration}),\ }%
  \bibfield{journal}{%
  \Doi{10.1103/PhysRevD.70.074513}{\bibinfo {journal} {Phys. Rev. D}}\ }%
  \textbf{\bibinfo {volume} {70}},\ \bibinfo {pages} {074513} (\bibinfo {year}
  {2004}),\ \Eprint{http://arxiv.org/abs/hep-lat/0402025}{arXiv:hep-lat/0402025
  [hep-lat]}%
  \bibAnnoteFile{NoStop}{Yamazaki:2004qb}%
\bibitem{Beane:2005rj}%
  \BibitemOpen
  \bibfield{author}{%
  \bibinfo {author} {\bibfnamefont{S.~R.}\ \bibnamefont{Beane}}, \bibinfo
  {author} {\bibfnamefont{P.~F.}\ \bibnamefont{Bedaque}}, \bibinfo {author}
  {\bibfnamefont{K.}~\bibnamefont{Orginos}},\ and\ \bibinfo {author}
  {\bibfnamefont{M.~J.}\ \bibnamefont{Savage}} (\bibinfo {collaboration}
  {NPLQCD Collaboration}),\ }%
  \bibfield{journal}{%
  \Doi{10.1103/PhysRevD.73.054503}{\bibinfo {journal} {Phys. Rev. D}}\ }%
  \textbf{\bibinfo {volume} {73}},\ \bibinfo {pages} {054503} (\bibinfo {year}
  {2006}),\ \Eprint{http://arxiv.org/abs/hep-lat/0506013}{arXiv:hep-lat/0506013
  [hep-lat]}%
  \bibAnnoteFile{NoStop}{Beane:2005rj}%
\bibitem{PhysRevD.77.094507}%
  \BibitemOpen
  \bibfield{author}{%
  \bibinfo {author} {\bibfnamefont{S.~R.}\ \bibnamefont{Beane}}, \bibinfo
  {author} {\bibfnamefont{T.~C.}\ \bibnamefont{Luu}}, \bibinfo {author}
  {\bibfnamefont{K.}~\bibnamefont{Orginos}}, \bibinfo {author}
  {\bibfnamefont{A.}~\bibnamefont{Parre\~no}}, \bibinfo {author}
  {\bibfnamefont{M.~J.}\ \bibnamefont{Savage}}, \bibinfo {author}
  {\bibfnamefont{A.}~\bibnamefont{Torok}},\ and\ \bibinfo {author}
  {\bibfnamefont{A.}~\bibnamefont{Walker-Loud}} (\bibinfo {collaboration}
  {NPLQCD Collaboration}),\ }%
  \bibfield{journal}{%
  \Doi{10.1103/PhysRevD.77.094507}{\bibinfo {journal} {Phys. Rev. D}}\ }%
  \textbf{\bibinfo {volume} {77}},\ \bibinfo {pages} {094507} (\bibinfo {month}
  {May}\ \bibinfo {year} {2008}),\
  \url{http://link.aps.org/doi/10.1103/PhysRevD.77.094507}%
  \bibAnnoteFile{NoStop}{PhysRevD.77.094507}%
\bibitem{PhysRevD.81.074506}%
  \BibitemOpen
  \bibfield{author}{%
  \bibinfo {author} {\bibfnamefont{A.}~\bibnamefont{Torok}}, \bibinfo {author}
  {\bibfnamefont{S.~R.}\ \bibnamefont{Beane}}, \bibinfo {author}
  {\bibfnamefont{W.}~\bibnamefont{Detmold}}, \bibinfo {author}
  {\bibfnamefont{T.~C.}\ \bibnamefont{Luu}}, \bibinfo {author}
  {\bibfnamefont{K.}~\bibnamefont{Orginos}}, \bibinfo {author}
  {\bibfnamefont{A.}~\bibnamefont{Parre\~no}}, \bibinfo {author}
  {\bibfnamefont{M.~J.}\ \bibnamefont{Savage}},\ and\ \bibinfo {author}
  {\bibfnamefont{A.}~\bibnamefont{Walker-Loud}} (\bibinfo {collaboration}
  {NPLQCD Collaboration}),\ }%
  \bibfield{journal}{%
  \Doi{10.1103/PhysRevD.81.074506}{\bibinfo {journal} {Phys. Rev. D}}\ }%
  \textbf{\bibinfo {volume} {81}},\ \bibinfo {pages} {074506} (\bibinfo {month}
  {Apr}\ \bibinfo {year} {2010}),\
  \url{http://link.aps.org/doi/10.1103/PhysRevD.81.074506}%
  \bibAnnoteFile{NoStop}{PhysRevD.81.074506}%
\bibitem{Feng2010268}%
  \BibitemOpen
  \bibfield{author}{%
  \bibinfo {author} {\bibfnamefont{X.}~\bibnamefont{Feng}}, \bibinfo {author}
  {\bibfnamefont{K.}~\bibnamefont{Jansen}},\ and\ \bibinfo {author}
  {\bibfnamefont{D.}~\bibnamefont{Renner}},\ }%
  \bibfield{journal}{%
  \Doi{http://dx.doi.org/10.1016/j.physletb.2010.01.018}{\bibinfo {journal}
  {Physics Letters B}}\ }%
  \textbf{\bibinfo {volume} {684}},\ \bibinfo {pages} {268 } (\bibinfo {year}
  {2010}),\ ISSN \bibinfo {issn} {0370-2693},\
  \url{http://www.sciencedirect.com/science/article/pii/S0370269310000602}%
  \bibAnnoteFile{NoStop}{Feng2010268}%
\bibitem{PhysRevD.86.034031}%
  \BibitemOpen
  \bibfield{author}{%
  \bibinfo {author} {\bibfnamefont{J.~J.}\ \bibnamefont{Dudek}}, \bibinfo
  {author} {\bibfnamefont{R.~G.}\ \bibnamefont{Edwards}},\ and\ \bibinfo
  {author} {\bibfnamefont{C.~E.}\ \bibnamefont{Thomas}} (\bibinfo
  {collaboration} {for the Hadron Spectrum Collaboration}),\ }%
  \bibfield{journal}{%
  \Doi{10.1103/PhysRevD.86.034031}{\bibinfo {journal} {Phys. Rev. D}}\ }%
  \textbf{\bibinfo {volume} {86}},\ \bibinfo {pages} {034031} (\bibinfo {month}
  {Aug}\ \bibinfo {year} {2012}),\
  \url{http://link.aps.org/doi/10.1103/PhysRevD.86.034031}%
  \bibAnnoteFile{NoStop}{PhysRevD.86.034031}%
\bibitem{PhysRevD.87.034505}%
  \BibitemOpen
  \bibfield{author}{%
  \bibinfo {author} {\bibfnamefont{J.~J.}\ \bibnamefont{Dudek}}, \bibinfo
  {author} {\bibfnamefont{R.~G.}\ \bibnamefont{Edwards}},\ and\ \bibinfo
  {author} {\bibfnamefont{C.~E.}\ \bibnamefont{Thomas}} (\bibinfo
  {collaboration} {for the Hadron Spectrum Collaboration}),\ }%
  \bibfield{journal}{%
  \Doi{10.1103/PhysRevD.87.034505}{\bibinfo {journal} {Phys. Rev. D}}\ }%
  \textbf{\bibinfo {volume} {87}},\ \bibinfo {pages} {034505} (\bibinfo {month}
  {Feb}\ \bibinfo {year} {2013}),\
  \url{http://link.aps.org/doi/10.1103/PhysRevD.87.034505}%
  \bibAnnoteFile{NoStop}{PhysRevD.87.034505}%
\bibitem{PhysRevD.87.054502}%
  \BibitemOpen
  \bibfield{author}{%
  \bibinfo {author} {\bibfnamefont{C.~B.}\ \bibnamefont{Lang}}\ and\ \bibinfo
  {author} {\bibfnamefont{V.}~\bibnamefont{Verduci}},\ }%
  \bibfield{journal}{%
  \Doi{10.1103/PhysRevD.87.054502}{\bibinfo {journal} {Phys. Rev. D}}\ }%
  \textbf{\bibinfo {volume} {87}},\ \bibinfo {pages} {054502} (\bibinfo {month}
  {Mar}\ \bibinfo {year} {2013}),\
  \url{http://link.aps.org/doi/10.1103/PhysRevD.87.054502}%
  \bibAnnoteFile{NoStop}{PhysRevD.87.054502}%
\bibitem{PhysRevLett.111.222001}%
  \BibitemOpen
  \bibfield{author}{%
  \bibinfo {author} {\bibfnamefont{D.}~\bibnamefont{Mohler}}, \bibinfo {author}
  {\bibfnamefont{C.~B.}\ \bibnamefont{Lang}}, \bibinfo {author}
  {\bibfnamefont{L.}~\bibnamefont{Leskovec}}, \bibinfo {author}
  {\bibfnamefont{S.}~\bibnamefont{Prelovsek}},\ and\ \bibinfo {author}
  {\bibfnamefont{R.~M.}\ \bibnamefont{Woloshyn}},\ }%
  \bibfield{journal}{%
  \Doi{10.1103/PhysRevLett.111.222001}{\bibinfo {journal} {Phys. Rev. Lett.}}\
  }%
  \textbf{\bibinfo {volume} {111}},\ \bibinfo {pages} {222001} (\bibinfo
  {month} {Nov}\ \bibinfo {year} {2013}),\
  \url{http://link.aps.org/doi/10.1103/PhysRevLett.111.222001}%
  \bibAnnoteFile{NoStop}{PhysRevLett.111.222001}%
\bibitem{PhysRevLett.111.192001}%
  \BibitemOpen
  \bibfield{author}{%
  \bibinfo {author} {\bibfnamefont{S.}~\bibnamefont{Prelovsek}}\ and\ \bibinfo
  {author} {\bibfnamefont{L.}~\bibnamefont{Leskovec}},\ }%
  \bibfield{journal}{%
  \Doi{10.1103/PhysRevLett.111.192001}{\bibinfo {journal} {Phys. Rev. Lett.}}\
  }%
  \textbf{\bibinfo {volume} {111}},\ \bibinfo {pages} {192001} (\bibinfo
  {month} {Nov}\ \bibinfo {year} {2013}),\
  \url{http://link.aps.org/doi/10.1103/PhysRevLett.111.192001}%
  \bibAnnoteFile{NoStop}{PhysRevLett.111.192001}%
\bibitem{Li:2003jn}%
  \BibitemOpen
  \bibfield{author}{%
  \bibinfo {author} {\bibfnamefont{X.}~\bibnamefont{Li}}\ and\ \bibinfo
  {author} {\bibfnamefont{C.}~\bibnamefont{Liu}},\ }%
  \bibfield{journal}{%
  \Doi{10.1016/j.physletb.2004.02.068}{\bibinfo {journal} {Phys. Lett. B}}\ }%
  \textbf{\bibinfo {volume} {587}},\ \bibinfo {pages} {100} (\bibinfo {year}
  {2004}),\
  \Eprint{http://arxiv.org/abs/hep-lat/0311035}{arXiv:hep-lat/0311035}%
  \bibAnnoteFile{NoStop}{Li:2003jn}%
\bibitem{Feng:2004ua}%
  \BibitemOpen
  \bibfield{author}{%
  \bibinfo {author} {\bibfnamefont{X.}~\bibnamefont{Feng}}, \bibinfo {author}
  {\bibfnamefont{X.}~\bibnamefont{Li}},\ and\ \bibinfo {author}
  {\bibfnamefont{C.}~\bibnamefont{Liu}},\ }%
  \bibfield{journal}{%
  \Doi{10.1103/PhysRevD.70.014505}{\bibinfo {journal} {Phys. Rev. D}}\ }%
  \textbf{\bibinfo {volume} {70}},\ \bibinfo {pages} {014505} (\bibinfo {year}
  {2004}),\
  \Eprint{http://arxiv.org/abs/hep-lat/0404001}{arXiv:hep-lat/0404001}%
  \bibAnnoteFile{NoStop}{Feng:2004ua}%
\bibitem{Rummukainen:1995vs}%
  \BibitemOpen
  \bibfield{author}{%
  \bibinfo {author} {\bibfnamefont{K.}~\bibnamefont{Rummukainen}}\ and\
  \bibinfo {author} {\bibfnamefont{S.~A.}\ \bibnamefont{Gottlieb}},\ }%
  \bibfield{journal}{%
  \Doi{10.1016/0550-3213(95)00313-H}{\bibinfo {journal} {Nucl. Phys. B}}\ }%
  \textbf{\bibinfo {volume} {450}},\ \bibinfo {pages} {397} (\bibinfo {year}
  {1995}),\ \Eprint{http://arxiv.org/abs/hep-lat/9503028}{arXiv:hep-lat/9503028
  [hep-lat]}%
  \bibAnnoteFile{NoStop}{Rummukainen:1995vs}%
\bibitem{XuFeng:2011}%
  \BibitemOpen
  \bibfield{author}{%
  \bibinfo {author} {\bibfnamefont{X.}~\bibnamefont{Feng}}, \bibinfo {author}
  {\bibfnamefont{K.}~\bibnamefont{Jansen}},\ and\ \bibinfo {author}
  {\bibfnamefont{D.~B.}\ \bibnamefont{Renner}} (\bibinfo {collaboration}
  {ETM}),\ }%
  \bibfield{booktitle}{%
  \emph{\bibinfo {booktitle} {{Proceedings, 28th International Symposium on
  Lattice field theory (Lattice 2010)}}},\ }%
  \bibfield{journal}{%
  \bibinfo {journal} {PoS LAT}\ }%
  \textbf{\bibinfo {volume} {2010}},\ \bibinfo {pages} {104} (\bibinfo {year}
  {2010}),\ \Eprint{http://arxiv.org/abs/1104.0058}{arXiv:1104.0058 [hep-lat]}%
  \bibAnnoteFile{NoStop}{XuFeng:2011}%
\bibitem{Davoudi:2011md}%
  \BibitemOpen
  \bibfield{author}{%
  \bibinfo {author} {\bibfnamefont{Z.}~\bibnamefont{Davoudi}}\ and\ \bibinfo
  {author} {\bibfnamefont{M.~J.}\ \bibnamefont{Savage}},\ }%
  \bibfield{journal}{%
  \Doi{10.1103/PhysRevD.84.114502}{\bibinfo {journal} {Phys. Rev. D}}\ }%
  \textbf{\bibinfo {volume} {84}},\ \bibinfo {pages} {114502} (\bibinfo {year}
  {2011}),\ \Eprint{http://arxiv.org/abs/1108.5371}{arXiv:1108.5371 [hep-lat]}%
  \bibAnnoteFile{NoStop}{Davoudi:2011md}%
\bibitem{ZiwenFu2012}%
  \BibitemOpen
  \bibfield{author}{%
  \bibinfo {author} {\bibfnamefont{Z.}~\bibnamefont{Fu}},\ }%
  \bibfield{journal}{%
  \Doi{10.1103/PhysRevD.85.074501}{\bibinfo {journal} {Phys. Rev. D}}\ }%
  \textbf{\bibinfo {volume} {85}},\ \bibinfo {pages} {074501} (\bibinfo {year}
  {2012}),\ \Eprint{http://arxiv.org/abs/1110.1422}{arXiv:1110.1422 [hep-lat]}%
  \bibAnnoteFile{NoStop}{ZiwenFu2012}%
\bibitem{Gockeler:2012yj}%
  \BibitemOpen
  \bibfield{author}{%
  \bibinfo {author} {\bibfnamefont{M.}~\bibnamefont{Gockeler}}, \bibinfo
  {author} {\bibfnamefont{R.}~\bibnamefont{Horsley}}, \bibinfo {author}
  {\bibfnamefont{M.}~\bibnamefont{Lage}}, \bibinfo {author}
  {\bibfnamefont{U.~G.}\ \bibnamefont{Meissner}}, \bibinfo {author}
  {\bibfnamefont{P.~E.~L.}\ \bibnamefont{Rakow}}, \bibinfo {author}
  {\bibfnamefont{A.}~\bibnamefont{Rusetsky}}, \bibinfo {author}
  {\bibfnamefont{G.}~\bibnamefont{Schierholz}},\ and\ \bibinfo {author}
  {\bibfnamefont{J.~M.}\ \bibnamefont{Zanotti}},\ }%
  \bibfield{journal}{%
  \Doi{10.1103/PhysRevD.86.094513}{\bibinfo {journal} {Phys. Rev. D}}\ }%
  \textbf{\bibinfo {volume} {86}},\ \bibinfo {pages} {094513} (\bibinfo {year}
  {2012}),\ \Eprint{http://arxiv.org/abs/1206.4141}{arXiv:1206.4141 [hep-lat]}%
  \bibAnnoteFile{NoStop}{Gockeler:2012yj}%
\bibitem{Bedaque:2004ax}%
  \BibitemOpen
  \bibfield{author}{%
  \bibinfo {author} {\bibfnamefont{P.~F.}\ \bibnamefont{Bedaque}}\ and\
  \bibinfo {author} {\bibfnamefont{J.-W.}\ \bibnamefont{Chen}},\ }%
  \bibfield{journal}{%
  \Doi{10.1016/j.physletb.2005.04.045}{\bibinfo {journal} {Phys. Lett. B}}\ }%
  \textbf{\bibinfo {volume} {616}},\ \bibinfo {pages} {208} (\bibinfo {year}
  {2005}),\ \Eprint{http://arxiv.org/abs/hep-lat/0412023}{arXiv:hep-lat/0412023
  [hep-lat]}%
  \bibAnnoteFile{NoStop}{Bedaque:2004ax}%
\bibitem{Bedaque:2004kc}%
  \BibitemOpen
  \bibfield{author}{%
  \bibinfo {author} {\bibfnamefont{P.~F.}\ \bibnamefont{Bedaque}},\ }%
  \bibfield{journal}{%
  \Doi{10.1016/j.physletb.2004.04.045}{\bibinfo {journal} {Phys. Lett. B}}\ }%
  \textbf{\bibinfo {volume} {593}},\ \bibinfo {pages} {82} (\bibinfo {year}
  {2004}),\ \Eprint{http://arxiv.org/abs/nucl-th/0402051}{arXiv:nucl-th/0402051
  [nucl-th]}%
  \bibAnnoteFile{NoStop}{Bedaque:2004kc}%
\bibitem{Sachrajda:2004mi}%
  \BibitemOpen
  \bibfield{author}{%
  \bibinfo {author} {\bibfnamefont{C.}~\bibnamefont{Sachrajda}}\ and\ \bibinfo
  {author} {\bibfnamefont{G.}~\bibnamefont{Villadoro}},\ }%
  \bibfield{journal}{%
  \Doi{10.1016/j.physletb.2005.01.033}{\bibinfo {journal} {Phys. Lett. B}}\ }%
  \textbf{\bibinfo {volume} {609}},\ \bibinfo {pages} {73} (\bibinfo {year}
  {2005}),\ \Eprint{http://arxiv.org/abs/hep-lat/0411033}{arXiv:hep-lat/0411033
  [hep-lat]}%
  \bibAnnoteFile{NoStop}{Sachrajda:2004mi}%
\bibitem{deDivitiis:2004kq}%
  \BibitemOpen
  \bibfield{author}{%
  \bibinfo {author} {\bibfnamefont{G.}~\bibnamefont{de~Divitiis}}, \bibinfo
  {author} {\bibfnamefont{R.}~\bibnamefont{Petronzio}},\ and\ \bibinfo {author}
  {\bibfnamefont{N.}~\bibnamefont{Tantalo}},\ }%
  \bibfield{journal}{%
  \Doi{10.1016/j.physletb.2004.06.035}{\bibinfo {journal} {Phys. Lett. B}}\ }%
  \textbf{\bibinfo {volume} {595}},\ \bibinfo {pages} {408} (\bibinfo {year}
  {2004}),\ \Eprint{http://arxiv.org/abs/hep-lat/0405002}{arXiv:hep-lat/0405002
  [hep-lat]}%
  \bibAnnoteFile{NoStop}{deDivitiis:2004kq}%
\bibitem{Beane:2006mx}%
  \BibitemOpen
  \bibfield{author}{%
  \bibinfo {author} {\bibfnamefont{S.}~\bibnamefont{Beane}}, \bibinfo {author}
  {\bibfnamefont{P.}~\bibnamefont{Bedaque}}, \bibinfo {author}
  {\bibfnamefont{K.}~\bibnamefont{Orginos}},\ and\ \bibinfo {author}
  {\bibfnamefont{M.}~\bibnamefont{Savage}},\ }%
  \bibfield{journal}{%
  \Doi{10.1103/PhysRevLett.97.012001}{\bibinfo {journal} {Phys. Rev. Lett.}}\
  }%
  \textbf{\bibinfo {volume} {97}},\ \bibinfo {pages} {012001} (\bibinfo {year}
  {2006}),\ \Eprint{http://arxiv.org/abs/hep-lat/0602010}{arXiv:hep-lat/0602010
  [hep-lat]}%
  \bibAnnoteFile{NoStop}{Beane:2006mx}%
\bibitem{Beane:2003da}%
  \BibitemOpen
  \bibfield{author}{%
  \bibinfo {author} {\bibfnamefont{S.}~\bibnamefont{Beane}}, \bibinfo {author}
  {\bibfnamefont{P.}~\bibnamefont{Bedaque}}, \bibinfo {author}
  {\bibfnamefont{A.}~\bibnamefont{Parreno}},\ and\ \bibinfo {author}
  {\bibfnamefont{M.}~\bibnamefont{Savage}},\ }%
  \bibfield{journal}{%
  \Doi{10.1016/j.physletb.2004.02.007}{\bibinfo {journal} {Phys. Lett. B}}\ }%
  \textbf{\bibinfo {volume} {585}},\ \bibinfo {pages} {106} (\bibinfo {year}
  {2004}),\ \Eprint{http://arxiv.org/abs/hep-lat/0312004}{arXiv:hep-lat/0312004
  [hep-lat]}%
  \bibAnnoteFile{NoStop}{Beane:2003da}%
\bibitem{Meng:2003gm}%
  \BibitemOpen
  \bibfield{author}{%
  \bibinfo {author} {\bibfnamefont{G.-w.}\ \bibnamefont{Meng}}, \bibinfo
  {author} {\bibfnamefont{C.}~\bibnamefont{Miao}}, \bibinfo {author}
  {\bibfnamefont{X.-n.}\ \bibnamefont{Du}},\ and\ \bibinfo {author}
  {\bibfnamefont{C.}~\bibnamefont{Liu}},\ }%
  \bibfield{journal}{%
  \Doi{10.1142/S0217751X04019627}{\bibinfo {journal} {Int. J. Mod. Phys. A}}\
  }%
  \textbf{\bibinfo {volume} {19}},\ \bibinfo {pages} {4401} (\bibinfo {year}
  {2004}),\ \Eprint{http://arxiv.org/abs/hep-lat/0309048}{arXiv:hep-lat/0309048
  [hep-lat]}%
  \bibAnnoteFile{NoStop}{Meng:2003gm}%
\bibitem{Bernard:2008ax}%
  \BibitemOpen
  \bibfield{author}{%
  \bibinfo {author} {\bibfnamefont{V.}~\bibnamefont{Bernard}}, \bibinfo
  {author} {\bibfnamefont{M.}~\bibnamefont{Lage}}, \bibinfo {author}
  {\bibfnamefont{U.-G.}\ \bibnamefont{Meissner}},\ and\ \bibinfo {author}
  {\bibfnamefont{A.}~\bibnamefont{Rusetsky}},\ }%
  \bibfield{journal}{%
  \Doi{10.1088/1126-6708/2008/08/024}{\bibinfo {journal} {JHEP}}\ }%
  \textbf{\bibinfo {volume} {0808}},\ \bibinfo {pages} {024} (\bibinfo {year}
  {2008}),\ \Eprint{http://arxiv.org/abs/0806.4495}{arXiv:0806.4495 [hep-lat]}%
  \bibAnnoteFile{NoStop}{Bernard:2008ax}%
\bibitem{Ishizuka:2009bx}%
  \BibitemOpen
  \bibfield{author}{%
  \bibinfo {author} {\bibfnamefont{N.}~\bibnamefont{Ishizuka}},\ }%
  \bibfield{booktitle}{%
  \emph{\bibinfo {booktitle} {{Proceedings, 27th International Symposium on
  Lattice field theory (Lattice 2009)}}},\ }%
  \bibfield{journal}{%
  \bibinfo {journal} {PoS LAT}\ }%
  \textbf{\bibinfo {volume} {2009}},\ \bibinfo {pages} {119} (\bibinfo {year}
  {2009}),\ \Eprint{http://arxiv.org/abs/0910.2772}{arXiv:0910.2772 [hep-lat]}%
  \bibAnnoteFile{NoStop}{Ishizuka:2009bx}%
\bibitem{Li:2007ey}%
  \BibitemOpen
  \bibfield{author}{%
  \bibinfo {author} {\bibfnamefont{X.}~\bibnamefont{Li}} \emph{et~al.}
  (\bibinfo {collaboration} {CLQCD}),\ }%
  \bibfield{journal}{%
  \Doi{10.1088/1126-6708/2007/06/053}{\bibinfo {journal} {JHEP}}\ }%
  \textbf{\bibinfo {volume} {06}},\ \bibinfo {pages} {053} (\bibinfo {year}
  {2007}),\
  \Eprint{http://arxiv.org/abs/hep-lat/0703015}{arXiv:hep-lat/0703015}%
  \bibAnnoteFile{NoStop}{Li:2007ey}%
\bibitem{Meng:2009qt}%
  \BibitemOpen
  \bibfield{author}{%
  \bibinfo {author} {\bibfnamefont{G.-Z.}\ \bibnamefont{Meng}} \emph{et~al.}
  (\bibinfo {collaboration} {CLQCD}),\ }%
  \bibfield{journal}{%
  \Doi{10.1103/PhysRevD.80.034503}{\bibinfo {journal} {Phys. Rev.}}\ }%
  \textbf{\bibinfo {volume} {D80}},\ \bibinfo {pages} {034503} (\bibinfo {year}
  {2009}),\ \Eprint{http://arxiv.org/abs/0905.0752}{arXiv:0905.0752 [hep-lat]}%
  \bibAnnoteFile{NoStop}{Meng:2009qt}%
\bibitem{Lee:2017igf}%
  \BibitemOpen
  \bibfield{author}{%
  \bibinfo {author} {\bibfnamefont{F.~X.}\ \bibnamefont{Lee}}\ and\ \bibinfo
  {author} {\bibfnamefont{A.}~\bibnamefont{Alexandru}}}%
   (\bibinfo {year} {2017}),\
  \Eprint{http://arxiv.org/abs/1706.00262}{arXiv:1706.00262 [hep-lat]}%
  \bibAnnoteFile{NoStop}{Lee:2017igf}%
\bibitem{Newton}%
  \BibitemOpen
  \bibfield{author}{%
  \bibinfo {author} {\bibfnamefont{R.}~\bibnamefont{Newton}},\ }%
  \emph{\bibinfo {title} {SCATTERING THEORY OF WAVES AND PARTICLES}}\ (\bibinfo
  {publisher} {McGraw-Hill},\ \bibinfo {address} {New York},\ \bibinfo {year}
  {1982})%
  \bibAnnoteFile{NoStop}{Newton}%
\bibitem{Blatt:1952zza}%
  \BibitemOpen
  \bibfield{author}{%
  \bibinfo {author} {\bibfnamefont{J.~M.}\ \bibnamefont{Blatt}}\ and\ \bibinfo
  {author} {\bibfnamefont{L.~C.}\ \bibnamefont{Biedenharn}},\ }%
  \bibfield{journal}{%
  \Doi{10.1103/PhysRev.86.399}{\bibinfo {journal} {Phys. Rev.}}\ }%
  \textbf{\bibinfo {volume} {86}},\ \bibinfo {pages} {399} (\bibinfo {year}
  {1952})%
  \bibAnnoteFile{NoStop}{Blatt:1952zza}%
\bibitem{Sasaki:2017ysy}%
  \BibitemOpen
  \bibfield{author}{%
  \bibinfo {author} {\bibfnamefont{K.}~\bibnamefont{Sasaki}} \emph{et~al.},\ }%
  \bibfield{booktitle}{%
  \emph{\bibinfo {booktitle} {{Proceedings, 34th International Symposium on
  Lattice Field Theory (Lattice 2016): Southampton, UK, July 24-30, 2016}}},\
  }%
  \bibfield{journal}{%
  \bibinfo {journal} {PoS}\ }%
  \textbf{\bibinfo {volume} {LATTICE2016}},\ \bibinfo {pages} {116} (\bibinfo
  {year} {2017}),\ \Eprint{http://arxiv.org/abs/1702.06241}{arXiv:1702.06241
  [hep-lat]}%
  \bibAnnoteFile{NoStop}{Sasaki:2017ysy}%
\bibitem{Nemura:2017bbw}%
  \BibitemOpen
  \bibfield{author}{%
  \bibinfo {author} {\bibfnamefont{H.}~\bibnamefont{Nemura}} \emph{et~al.},\ }%
  \bibfield{booktitle}{%
  \emph{\bibinfo {booktitle} {{Proceedings, 34th International Symposium on
  Lattice Field Theory (Lattice 2016): Southampton, UK, July 24-30, 2016}}},\
  }%
  \bibfield{journal}{%
  \bibinfo {journal} {PoS}\ }%
  \textbf{\bibinfo {volume} {LATTICE2016}},\ \bibinfo {pages} {101} (\bibinfo
  {year} {2017}),\ \Eprint{http://arxiv.org/abs/1702.00734}{arXiv:1702.00734
  [hep-lat]}%
  \bibAnnoteFile{NoStop}{Nemura:2017bbw}%
\bibitem{Doi:2017cfx}%
  \BibitemOpen
  \bibfield{author}{%
  \bibinfo {author} {\bibfnamefont{T.}~\bibnamefont{Doi}} \emph{et~al.},\ }%
  \bibfield{booktitle}{%
  \emph{\bibinfo {booktitle} {{Proceedings, 34th International Symposium on
  Lattice Field Theory (Lattice 2016): Southampton, UK, July 24-30, 2016}}},\
  }%
  \bibfield{journal}{%
  \bibinfo {journal} {PoS}\ }%
  \textbf{\bibinfo {volume} {LATTICE2016}},\ \bibinfo {pages} {110} (\bibinfo
  {year} {2017}),\ \Eprint{http://arxiv.org/abs/1702.01600}{arXiv:1702.01600
  [hep-lat]}%
  \bibAnnoteFile{NoStop}{Doi:2017cfx}%
\bibitem{Ishii:2017xud}%
  \BibitemOpen
  \bibfield{author}{%
  \bibinfo {author} {\bibfnamefont{N.}~\bibnamefont{Ishii}} \emph{et~al.},\ }%
  \bibfield{booktitle}{%
  \emph{\bibinfo {booktitle} {{Proceedings, 34th International Symposium on
  Lattice Field Theory (Lattice 2016): Southampton, UK, July 24-30, 2016}}},\
  }%
  \bibfield{journal}{%
  \bibinfo {journal} {PoS}\ }%
  \textbf{\bibinfo {volume} {LATTICE2016}},\ \bibinfo {pages} {127} (\bibinfo
  {year} {2017}),\ \Eprint{http://arxiv.org/abs/1702.03495}{arXiv:1702.03495
  [hep-lat]}%
  \bibAnnoteFile{NoStop}{Ishii:2017xud}%
\end{thebibliography}%
\input Phase_arxiv.bbl

\end{document}